\documentclass[10pt,conference]{IEEEtran}
\IEEEoverridecommandlockouts
\usepackage{cite}
\usepackage{amsmath,amssymb,amsfonts}
\usepackage{graphicx}
\usepackage{textcomp}
\usepackage{xcolor}
\usepackage{algpseudocode}
\usepackage{algorithm,algorithmicx,algcompatible}
\usepackage{braket}
\usepackage{booktabs,multirow}

\def\BibTeX{{\rm B\kern-.05em{\sc i\kern-.025em b}\kern-.08em
    T\kern-.1667em\lower.7ex\hbox{E}\kern-.125emX}}

\usepackage{adjustbox}
\usepackage{tikz}
\usetikzlibrary{quantikz}
\usetikzlibrary{external}
\newcommand{\dicke}[2]{\ket{\smash{D_{#2}^{#1}}}}
\newcommand{\dsu}[1]{\textit{DSU}(#1)}
\newcommand{\scs}[1]{\textit{SCS}(#1)}
\newcommand{\ryangle}[2]{2\arccos(\sqrt{#1/#2})}
\newcommand{\rygate}[2]{\gate{R_y(\ryangle{#1}{#2})}}

\usepackage[colorlinks=true]{hyperref}

\usepackage[switch]{lineno}
\newcommand*\patchAMSlineno[1]{
	\expandafter\let\csname old#1\expandafter\endcsname\csname #1\endcsname
	\expandafter\let\csname oldend#1\expandafter\endcsname\csname end#1\endcsname
	\renewenvironment{#1}
		{\linenomath\csname old#1\endcsname}
		{\csname oldend#1\endcsname\endlinenomath}
}
\AtBeginDocument{
	\patchAMSlineno{equation}
	\patchAMSlineno{equation*}
	\patchAMSlineno{align}
	\patchAMSlineno{align*}
}

\begin{document}


\title{Optimized Telecloning Circuits: Theory and Practice of Nine NISQ Clones  \\
{}
\thanks{LA-UR-22-30899; 
 this work was supported by the U.S. Department of Energy through the Los Alamos National Laboratory. Los Alamos National Laboratory is operated by Triad National Security, LLC, for the National Nuclear Security Administration of U.S. Department of Energy (Contract No. 89233218CNA000001). 
We acknowledge the use of IBM Quantum services for this work. The views expressed are those of the authors, and do not reflect the official policy or position of IBM or the IBM Quantum team.
This research used resources of the Oak Ridge Leadership Computing Facility, which is a DOE Office of Science User Facility supported under Contract DE-AC05-00OR22725.
}
}

\author{
    \IEEEauthorblockN{Elijah Pelofske\IEEEauthorrefmark{1}\IEEEauthorrefmark{2}, 
    Andreas Bärtschi\IEEEauthorrefmark{1}\IEEEauthorrefmark{2}, 
    Stephan Eidenbenz\IEEEauthorrefmark{2}}
    \IEEEauthorblockA{\IEEEauthorrefmark{2}
    \textit{CCS-3 Information Sciences, Los Alamos National Laboratory},
	Los Alamos, NM 87544, USA
	}
    \IEEEauthorblockA{\IEEEauthorrefmark{1}
    Corresponding author: epelofske@lanl.gov
	}
}

\maketitle

\begin{abstract}
Although perfect copying of an unknown quantum state is not possible, approximate cloning is possible in quantum mechanics. Quantum telecloning is a variant of approximate quantum cloning which uses quantum teleportation to allow for the use of classical communication to create physically separate clones of a quantum state. We present results of a of $1 \rightarrow 9$ universal, symmetric, optimal quantum telecloning implementation on a cloud accessible quantum computer - the Quantinuum H1-1 device. The H1-1 device allows direct creation of the telecloning protocol due to real time classical if-statements that are conditional on the mid-circuit measurement outcome of a Bell measurement. In this implementation, we also provide an improvement over previous work for the circuit model description of quantum telecloning, which reduces the required gate depth and gate count for an all-to-all connectivity. The demonstration of creating $9$ approximate clones on a quantum processor is the largest number of clones that has been generated, telecloning or otherwise. 
\end{abstract}


\section{Introduction}
\label{sec:introduction}
Due to the no cloning theorem, perfect copies of an unknown quantum state can not be made \cite{wootters1982single}. However, approximate copies of an unknown state can be made \cite{buvzek1996quantum}---this process is referred to as \emph{quantum cloning}. There are a large number of variants of quantum cloning algorithms, and therefore it is helpful to classify these algorithms using different characteristics. \emph{Symmetric} quantum cloning means that all generated clones are identical and therefore have the same fidelity, whereas \emph{asymmetric} quantum cloning is where the clones can be different. \emph{Universal} quantum cloning is state independent, i.e., the clone quality is not dependent on the state being cloned. In state dependent versions of quantum cloning, on the other hand, the clone quality is dependent on the state which is cloned \cite{fan2001quantum}.

The optimal theoretical approximate clone fidelity limit for symmetric universal quantum cloning can be exactly computed \cite{scarani2005quantum, PhysRevA.59.156}, and is given in Eq. \eqref{eq:theoretical-fidelity}. A fidelity of $1$ indicates that the two quantum states being compared are identical, whereas a fidelity of $0$ indicates that the two quantum states are orthogonal,
while a maximally mixed (e.g. completely noisy) 1-qubit state results in a fidelity of $0.5$. 
\begin{equation}
    F_{N \rightarrow M} = \frac{ MN + M + N }{ M(N+2) } 
    \label{eq:theoretical-fidelity}
\end{equation}
A quantum cloning process which produces clones that can achieve this bound is referred to as \emph{optimal} \cite{gisin1997optimal, werner1998optimal}. 
Quantum telecloning is a combination of quantum teleportation \cite{bennett1993teleporting} and optimal quantum cloning \cite{PhysRevA.59.156}. Quantum telecloning allows the distribution of quantum information, specifically approximate clones of an unknown quantum state, to be distributed to different parties \cite{fan2014quantum}. In particular, the usage of classical communication of the measurement of a Bell state allows different parties (which could be spatially separated) to conditionally apply quantum operations to their qubits in order to generate optimal clones of the unknown quantum state. Algorithm \ref{algorithm:telecloning} details the telecloning process. Quantum cloning is of particular interest for quantum information processing and quantum networking \cite{Bennett_2014, PhysRevLett.88.127902, PhysRevA.60.2764, scarani2005quantum}.

Designing quantum telecloning circuits to run on Noise Intermediate Scale Quantum (NISQ) \cite{Preskill_2018} equipment is an open challenge because of the need for optimized, low-depth circuits due to noise and decoherence on NISQ devices. 
Using Dicke state preparation improvements \cite{baertschi2019deterministic,9774323,baertschi2022shortdepth,PhysRev.93.99,aulbach2010maximally,G_hne_2008}, the authors of \cite{pelofske2022telecloning} build an explicit circuit model description for implementing telecloning on NISQ devices. Our contributions are a significantly improved telecloning circuit over previous work that we show to be useful on a NISQ device by generating 9 clones.
More precisely, we:
\begin{enumerate}
    \item Provide an optimized circuit model algorithm of $1 \rightarrow M$ optimal, universal, symmetric telecloning for an all-to-all connectivity. 
    \item Report experimental results of $1 \rightarrow 9$ quantum telecloning on the Quantinuum H1-1 device, which is the largest experimental demonstration (in terms of the number of clones) of optimal universal symmetric clones generated on a NISQ computer.
\end{enumerate}

\begin{algorithm}[t!]
\caption{Quantum $1\rightarrow M$ Telecloning Protocol}
\begin{algorithmic}[1]
\Statex \hspace*{-2.5ex}\textbf{State Preparation:}
\State A message qubit $q_m$ is prepared by a sender
\State A quantum telecloning state $TC$ is constructed with
\Statex -~$(M-1)$ ancilla qubits $A$, $1$ Port qubit $P$, and
\Statex -~$M$ clone qubits $C$ (sent to the receivers).
\Statex \hspace*{-2.5ex}\textbf{Teleportation:}
\State A bell measurement is made between $q_m$ and $P$, and the results are communicated classically to the clone holders.
\State The clone holders use the result of the bell measurement to decide whether to apply $X$- and/or $Z$-gates to the clone qubits in order to construct the approximate clones:
\Statex - $\Phi^+$: apply nothing
\hspace*{20pt} - $\Phi^-$: apply $Z$-gate
\Statex - $\Psi^+$: apply $X$-gate
\hspace*{21pt} - $\Psi^-$: apply $X$- then $Z$-gate
\Statex \hspace*{-2.5ex}\textbf{Result:}
\State $M$ approximate clones of $q_m$ have been generated with theoretical maximal fidelity described by Eq.~\eqref{eq:theoretical-fidelity}.
\end{algorithmic}
\label{algorithm:telecloning}
\end{algorithm}

We find that (i) our $1 \rightarrow 9$ quantum telecloning circuit has 211 two-qubit gates; (ii) the fidelities we achieve vary across the nine individual clones and also across the four different cloned states that we tested, but generally range from $0.55$ to $0.67$ (theoretical optimum is $\frac{19}{27} \approx 0.7037$, see \eqref{eq:theoretical-fidelity}) with an average of 0.59 across all experiments, which is comparable to the fidelities achieved by e.g., the IBM Q devices for much smaller $1 \rightarrow 3$ telecloning circuits with ancilla \cite{pelofske2022telecloning}; (iii) beyond the standard use cases for telecloning, we find telecloning to be a good benchmark to test NISQ devices as a more detailed analysis reveals asymmetries in the results that are due to hard- and control-software details of the H1-1 device.

Earlier works have looked at telecloning with either two or three clones and mostly on custom-built devices, whereas our results include nine clones achieved on a fairly standard general-purpose, cloud-accessible quantum computing NISQ device: \cite{PhysRevLett.96.060504} reports fidelities of 0.58 for two teleclones; \cite{PhysRevA.104.032419} describes a device for 3 teleclones achieving fidelity 0.64 for two and 0.49 for the third clone; 
\cite{Chiuri_2012} provides a high-fidelity implementation of three teleclones on a custom-built photonics device; 
finally, \cite{pelofske2022telecloning} runs a large number of two and three telecloning experiments on IBM Q quantum computers as well as on Quantinuum's H1-1 device and reporting fidelities of 0.82 and 0.76 for two and three teleclones, respectively, on Quantinuum H1-2 and significantly lower fidelities for the IBM Q devices with additional parameter studies of circuits with and without ancilla qubits and error mitigation techniques. See \cite{fan2014quantum} for a review on telecloning and \cite{scarani2005quantum} for a review on quantum cloning. 

Figures are generated using a combination of Qiskit~\cite{Qiskit}, Matplotlib~\cite{thomas_a_caswell_2021_5194481, Hunter:2007}, QuTiP~\cite{JOHANSSON20121760, JOHANSSON20131234} and mayavi~\cite{ramachandran2011mayavi}. All data, code, and extra figures are available on a public Github repository\footnote{\url{https://github.com/lanl/Quantum-Telecloning}}. 

\begin{figure}[h]
    \centering
    \vspace{-1ex}
    \includegraphics[width=0.24\textwidth]{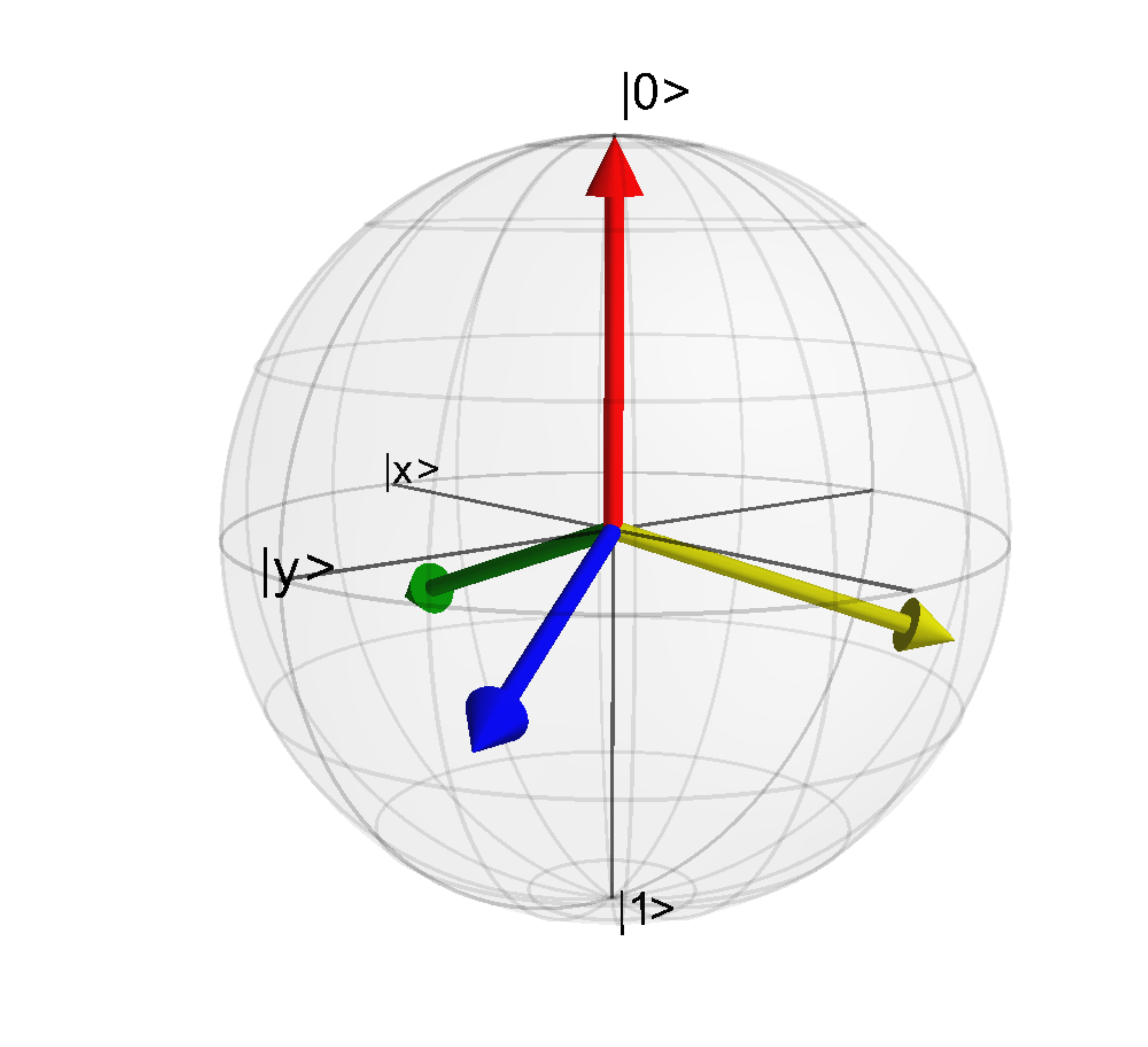}\hfill%
    \includegraphics[width=0.24\textwidth]{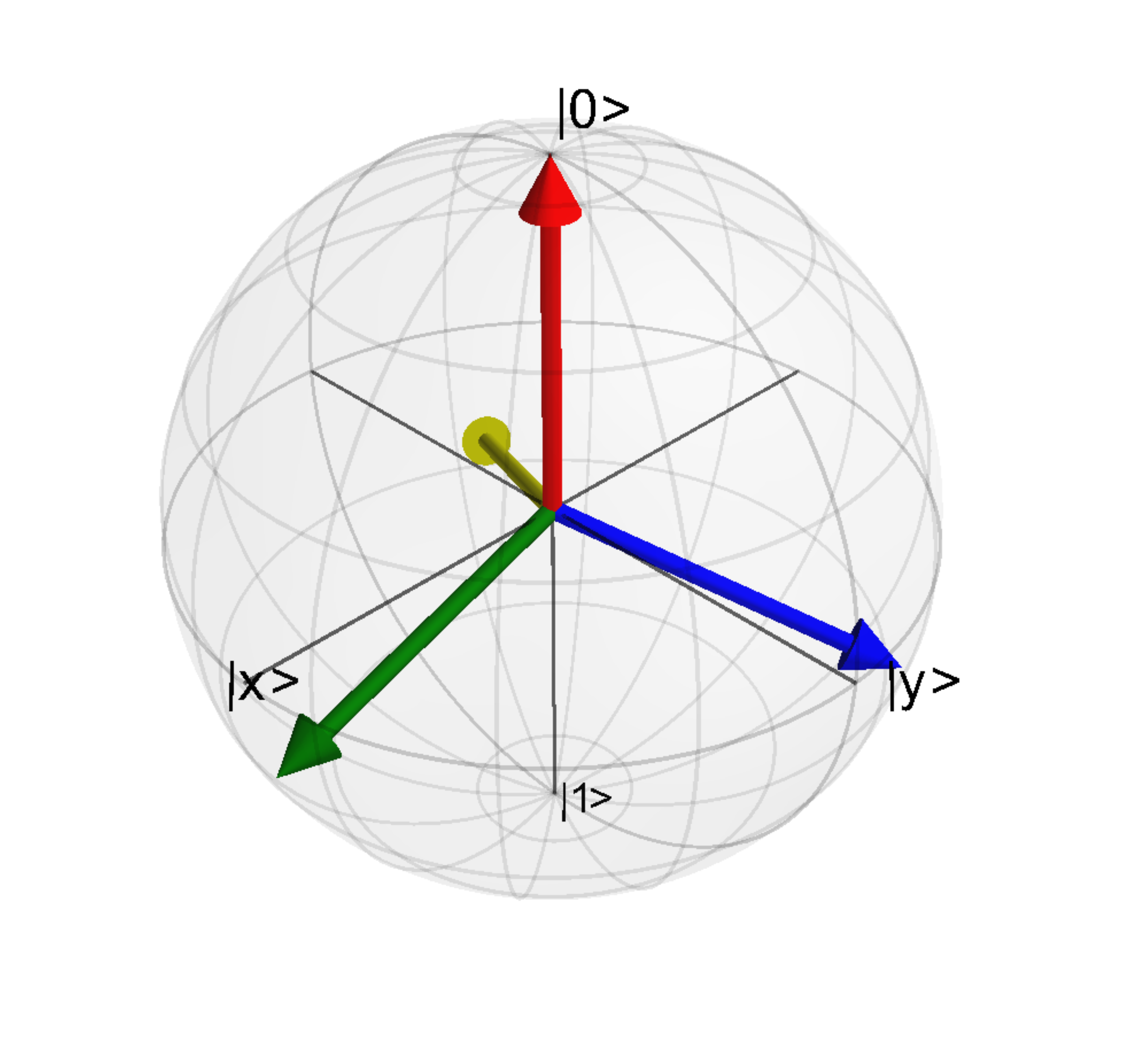}%
    \caption{Two different views of Bloch sphere representations of the four Tetrahedral basis states which we want to create approximate teleclones of.%
    }
    \label{fig:tetrahedral}
\end{figure}
%

\begin{figure*}[t!]
	\centering
	\begin{adjustbox}{width=\linewidth}
		\begin{quantikz}[row sep={24pt,between origins},execute at end picture={
					\node[fit=(\tikzcdmatrixname-2-2)(\tikzcdmatrixname-2-3),draw,thick,fill=white,inner xsep=24pt,inner ysep=10pt,xshift=0pt,yshift=0pt,label={[yshift=-20pt]$R_y(\ryangle{1}{10})$}] {};										
					\node[fit=(\tikzcdmatrixname-2-14)(\tikzcdmatrixname-6-15),draw,dashed,thick,rounded corners,inner xsep=14pt,inner ysep=7pt,xshift=10pt,yshift=-4.5pt,label={[yshift=-2pt]$\dsu{9}$}] {};					
					\node[fit=(\tikzcdmatrixname-7-14)(\tikzcdmatrixname-11-15),draw,thick,fill=white,inner xsep=19pt,inner ysep=10pt,xshift=0pt,yshift=0pt,label={[yshift=-70pt]$\dsu{9}$}] {};					
			}]
			\lstick{$q_m\colon\ket{0}$}	& \gate{R_y(\psi)}	& \gate{R_z(\phi)}	& \qw			& \qw					& \qw		& \qw		& \qw		& \qw		& \qw		& \qw		& \qw\slice{}	& \qw	& \qw				& \qw				& \qw\slice{}		& \ctrl{5}	& \gate{H}		& \meter{}		& \cw			& \cwbend{10}		&		\\
			\lstick{$A_1\colon \ket{0}$}	& \qw			& \qw			& \ctrl{1}		& \qw					& \qw		& \qw		& \ctrl{5}	& \qw		& \qw		& \qw		& \qw		& \qw	& \gate[5,nwires={3}]{\scs{9}}	& \gate[4,nwires={3}]{\dsu{8}}	& \trash{}		& 		& 			& 			&			&			&		\\ 
			\lstick{$A_2\colon \ket{0}$}	& \qw			& \qw			& \rygate{1}{9}		& \ctrl{1}				& \qw		& \qw		& \qw		& \ctrl{5}	& \qw		& \qw		& \qw		& \qw	& \qw				& \qw				& \trash{}		& 		& 			& 			&			&			& 		\\
			\lstick{$\ldots$}		& 			& 			& 			& \hspace*{8ex}\ldots\hspace*{-8ex}	& \vqw{1}	& 		& 		& 		& \ldots	& 		& 		& 	& 				& 				& \text{discard}	& 		& 			& 			&			&			& 		\\
			\lstick{$A_8\colon \ket{0}$}	& \qw			& \qw			& \qw			& \qw					& \rygate{1}{3}	& \ctrl{1}	& \qw		& \qw		& \qw		& \ctrl{5}	& \qw		& \qw	& \qw				& \qw				& \trash{}		& 		& 			& 			&			&			& 		\\
			\lstick{$P\ \colon \ket{0}$}	& \qw			& \qw			& \qw			& \qw					& \qw		& \rygate{1}{2}	& \qw		& \qw		& \qw		& \qw		& \ctrl{5}	& \qw	& \qw				& \qw				& \qw			& \targ{}	& \qw			& \meter{}		& \cwbend{5}		&			& 		\\
			\lstick{$C_1\colon \ket{0}$}	& \qw			& \qw			& \qw			& \qw					& \qw		& \qw		& \targ{}	& \qw		& \qw		& \qw		& \qw		& \qw	& \qw				& \qw				& \qw			& \qw		& \qw			& \qw			& \gate{X}		& \gate{Z}		& \qw		\\
			\lstick{$C_2\colon \ket{0}$}	& \qw			& \qw			& \qw			& \qw					& \qw		& \qw		& \qw		& \targ{}	& \qw		& \qw		& \qw		& \qw	& \qw				& \qw				& \qw			& \qw		& \qw			& \qw			& \gate{X}		& \gate{Z}		& \qw		\\
			\lstick{$\ldots$}		& 			& 			& 			& 					& 		& 		& 		& 		& \ldots	& 		& 		& 	& 				& 				& 			& 		& 			& 			& \gate[nwires={1}]{X}	& \gate[nwires={1}]{Z}	& \hspace*{-2ex}\ldots	\\
			\lstick{$C_8\colon \ket{0}$}	& \qw			& \qw			& \qw			& \qw					& \qw		& \qw		& \qw		& \qw		& \qw		& \targ{}	& \qw		& \qw	& \qw				& \qw				& \qw			& \qw		& \qw			& \qw			& \gate{X}		& \gate{Z}		& \qw		\\
			\lstick{$C_9\colon \ket{0}$}	& \qw			& \qw			& \qw			& \qw					& \qw		& \qw		& \qw		& \qw		& \qw		& \qw		& \targ{}	& \qw	& \qw				& \qw				& \qw			& \qw		& \qw			& \qw			& \gate{X}		& \gate{Z}		& \qw				
		\end{quantikz}
	\end{adjustbox}

	\caption{Quantum telecloning circuit for $M=9$ clones. We use little-endian ket notation (top-to-bottom wires correspond to left-to-right bitstrings):\newline
		\emph{(left)} First prepare the message qubit on the top wire and the state $\tfrac{1}{\smash{\surd{10}}}\sum_{i=0}^9 \ket{1^i0^{9-i}}\ket{1^i0^{9-i}}$ on 8 ancilla, 1 port and 9 clone qubits on the wires below.\newline
		\emph{(middle)} Symmetrize each of the parts $A^8P$ and $C^9$ by a Dicke State Unitary $\dsu{9}$ resulting in the telecloning state $\tfrac{1}{\smash{\surd{10}}} \sum_{i=0}^9 \dicke{9}{i}_{A^{8}P}\dicke{9}{i}_{C^9}$.
		Since the ancilla qubits are discarded, it is equally sufficient to only implement a Split \& Cyclic Shift unitary $\scs{9}$ on $A^8P$ and drop the consecutive $\dsu{8}$.
		This instead results in the corresponding telecloning state  $\tfrac{1}{\smash{\surd{10}}} \sum_{i=0}^9 (\surd\tfrac{9-i}{9}\ket{1^i0^{8-i}}_{A^{8}}\ket{0}_P + \surd\tfrac{i}{9}\ket{1^{i-1}0^{9-i}}_{A^{8}}\ket{1}_P)\dicke{9}{i}_{C^9}$.\newline
		\emph{(right)} Perform a Bell measurement on the message and port qubits, followed by classical communication to and local Pauli operations on the clone qubits.%
	}
	\label{fig:main-circuit}
\end{figure*}
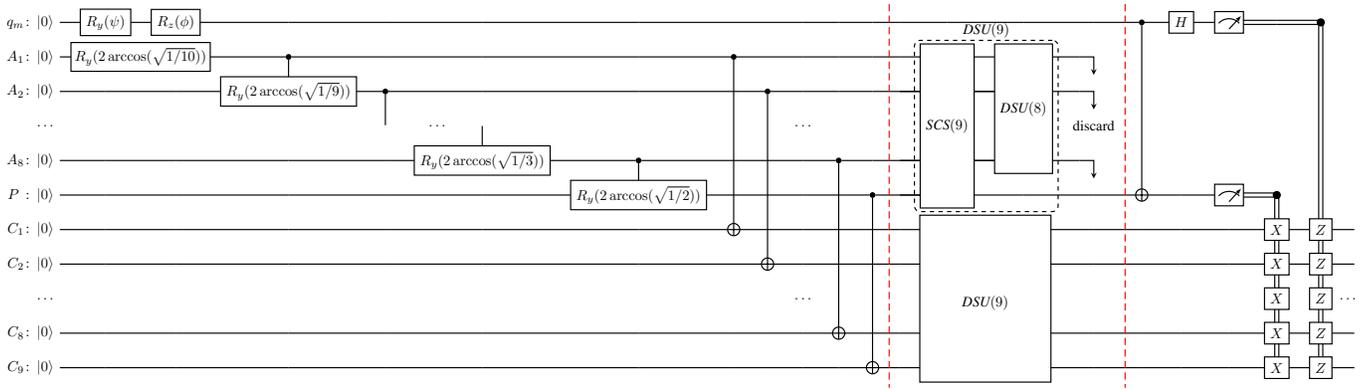

\begin{figure*}[h!]
    \centering
    \includegraphics[width=0.9\textwidth]{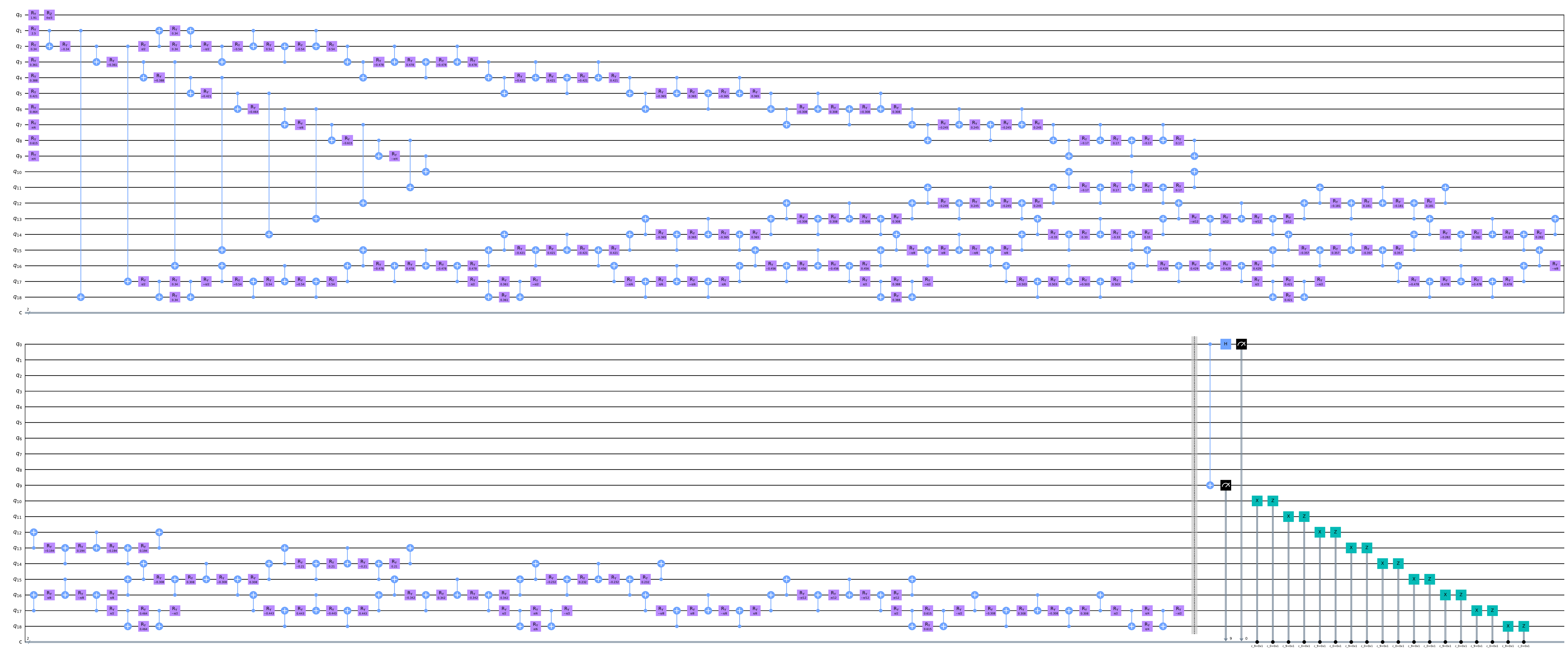}
    \caption{A detailed Qiskit based gate level description of Figure \ref{fig:main-circuit}. The circuit depth is $182$ and a total of 19 qubits are used. The circuit is comprised of $211$ CNOT gates, $194$ ry gates, $1$ Hadamard and $1$ rz gate, and $9$ X and Z Pauli gates which are conditioned using classical if-statements at the end of the circuit. Qubit register 0 is where the message qubit state is prepared with a single $ry$ and a single $rz$ gate. Notice that qubits 1 through 8 are ancilla qubits and are discarded at the end of the computation (meaning their state is not measured), qubit 9 is the Port qubit, and qubits 10 through 18 are the approximate symmetric clone qubits. A barrier is inserted into the circuit just before the Bell measurement is made in order to ensure that the Bell measurement (and subsequent if-statements) occur after the telecloning state has been prepared. Note that the clone qubit wire order is reverse of the order shown in Figure \ref{fig:main-circuit}. }
    \label{fig:M9_circuit}
\end{figure*}

\section{Methods}
\label{sec:methods}

\subsection{Algorithm improvements}
\label{sec:methods_algorithm}
Figure \ref{fig:main-circuit}, illustrated with quantikz \cite{quantikz}, details the all-to-all telecloning circuit implementation. This circuit construction reduces the number of two qubit gates (CNOTs) required by $\approx 33\%$ compared to previous work~\cite{pelofske2022telecloning}, while keeping the circuit depth the same. Our circuit construction works for any $M$; we illustrate it for $M=9$, the largest such circuit that can be executed on current Quantinuum H1 devices.

For $1\rightarrow M$ quantum telecloning, one generates a message qubit $q_m$ and a \emph{telecloning state} $A^{(M-1)}PC^M$ on $M-1$ ancilla, 1 port, and $M$ clone qubits of the form~\cite{PhysRevA.59.156}
\begin{align}
    \ket{A^{M-1}PC^M} = \tfrac{1}{\sqrt{M+1}} \sum\nolimits_{i=0}^M \dicke{M}{i}_{A^{M-1}P} \dicke{M}{i}_{C^M},
\end{align}
where $\dicke{M}{i}$ denotes the uniform superposition over all $M$-qubit states of Hamming weight $i$ with real amplitudes, e.g., $\dicke{3}{1} = \left[ \ket{001} + \ket{010} + \ket{100} \right] / \sqrt{3}$.
After creation of this state, the ancilla qubits \emph{are discarded}.

It is known~\cite{pelofske2022telecloning} that such a telecloning state can be created using Dicke state unitaries $\dsu{M}$, based themselves on Split \& Cyclic Shift unitaries $\scs{m}$~\cite{baertschi2019deterministic}, defined by:
\begin{align*}
    \dsu{M}\colon & \ket{1^i 0^{M-i}} \mapsto \dicke{M}{i},   \\
    \scs{m}\colon & \ket{1^i 0^{m-i}} \mapsto \surd \tfrac{m-i}{m} \ket{1^i 0^{m-i}} + \surd \tfrac{i}{m} \ket{1^{i-1} 0^{m-i}1}.
\end{align*}
We make two observations, cf. Figure~\ref{fig:main-circuit}:
\textbf{(i)} the telecloning state can be created by applying two Dicke state unitaries $\dsu{M}$ in parallel to the (easily creatable) input state $\tfrac{1}{\sqrt{M+1}} \sum\nolimits_{i=0}^M \ket{1^i0^{M-i}}_{A^{M-1}P} \ket{1^i0^{M-i}}_{C^M}$, and\linebreak
\textbf{(ii)} Dicke state unitaries are made up of Split \& Cyclic Shift unitaries: 
$\dsu{M} = \prod_{m=1}^M \scs{m}\otimes \textit{Id}^{M-m}$.
We make use of these observations combined with the fact that ancilla qubits are discarded:
When considering the application of $\dsu{M}$ to the ancilla and port qubits, only the first $\scs{M}$ unitary involves the port qubit, the following $\scs{m}$ unitaries comprising a $\dsu{M-1}$ unitary act only on the to be discarded ancilla qubits. 
These unitaries can thus \emph{be removed} without consequence to the telecloning protocol, resulting in a reduction from $330$ to $211$ CNOT gates for $1\rightarrow 9$ telecloning.

Finally, after distribution of the clones to the receivers, the standard teleportation protocol~\cite{bennett1993teleporting} is executed. In this, the sender measures its $\ket{\text{message,port}}$ qubits in the Bell basis. The result is communicated classically to the receivers, which -- depending on the measurement -- adjust their clones with/without local Pauli-$X$ and/or $Z$ gates, see Algorithm~\ref{algorithm:telecloning}.

\subsection{Experimental telecloning implementation}
\label{sec:methods_experimental_setup}
The improved all-to-all circuit descriptions for the telecloning protocol are implemented on the Quantinuum H1-1 device, which has an all-to-all two qubit gate connectivity \cite{Pino_2021}, and which allows a direct implementation of classical if-statements based on the result of the mid-circuit Bell measurement. 
Due to speed limitations of the H1-1 trapped-ion architecture \cite{Pino_2021}, the circuit size and number of samples we can compute at reasonable cost is more constrained than on superconducting circuit based quantum computers (see \cite{pelofske2022telecloning}) resulting in a relatively sparse parameter exploration that still reveals general details about the performance of the H1-1 system. We test the telecloning circuit on four well-defined pure quantum states which form the tetrahedral basis states on the Bloch sphere \cite{single-qubit-state-tomography} at angles $2acos(\sqrt{1/3}) \approx 109.5$ degrees from each other. Bloch sphere drawings of these states are given in Figure \ref{fig:tetrahedral}. These states are constructed on the message qubit using an angle parameterized $ry$ gate and an angle parameterized $rz$ gate. This set of states was chosen because it is a small but reasonable test set of pure quantum states distributed over the Bloch sphere. 

\begin{figure*}[h!]
    \centering
    \includegraphics[width=0.45\textwidth]{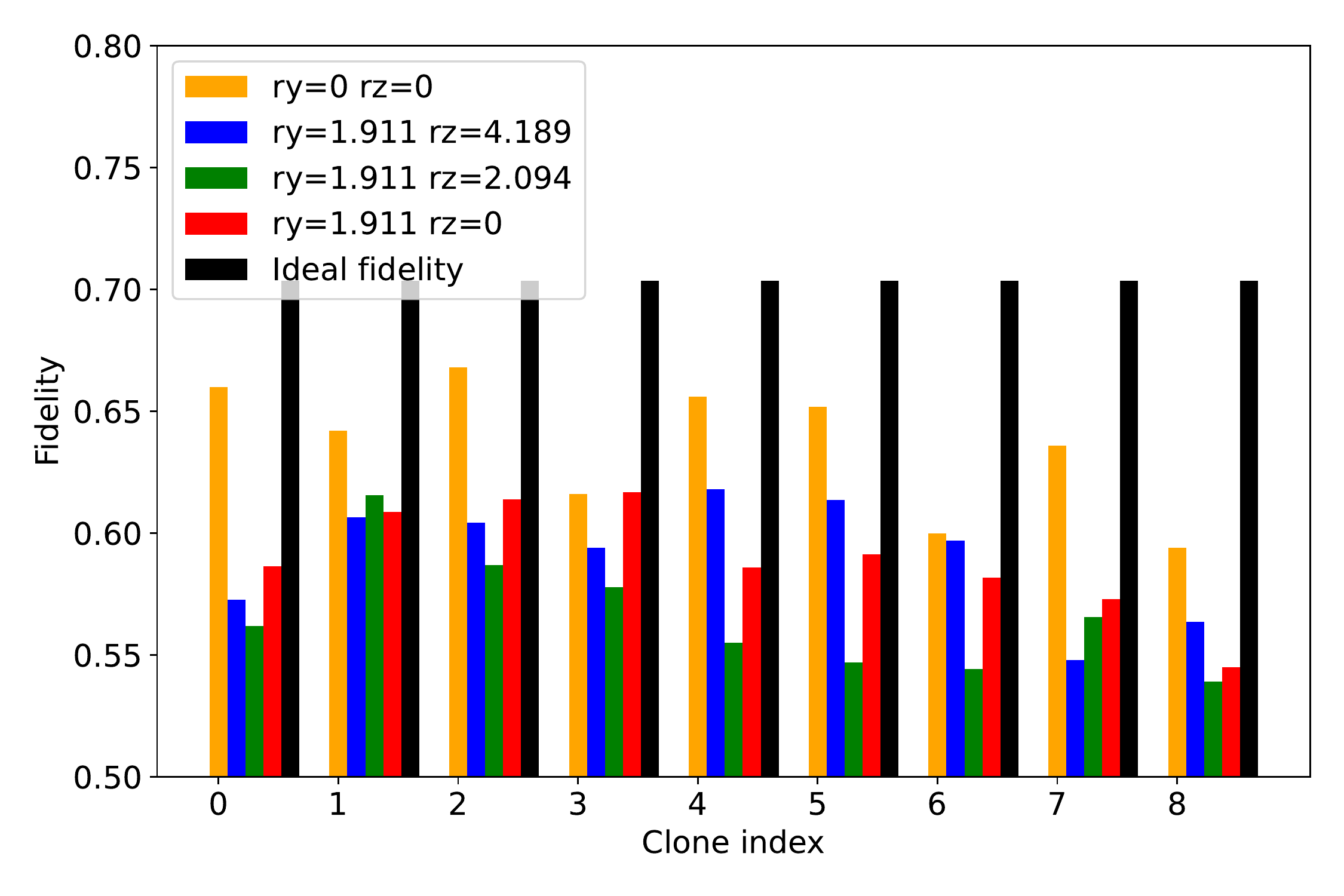}
    \includegraphics[width=0.45\textwidth]{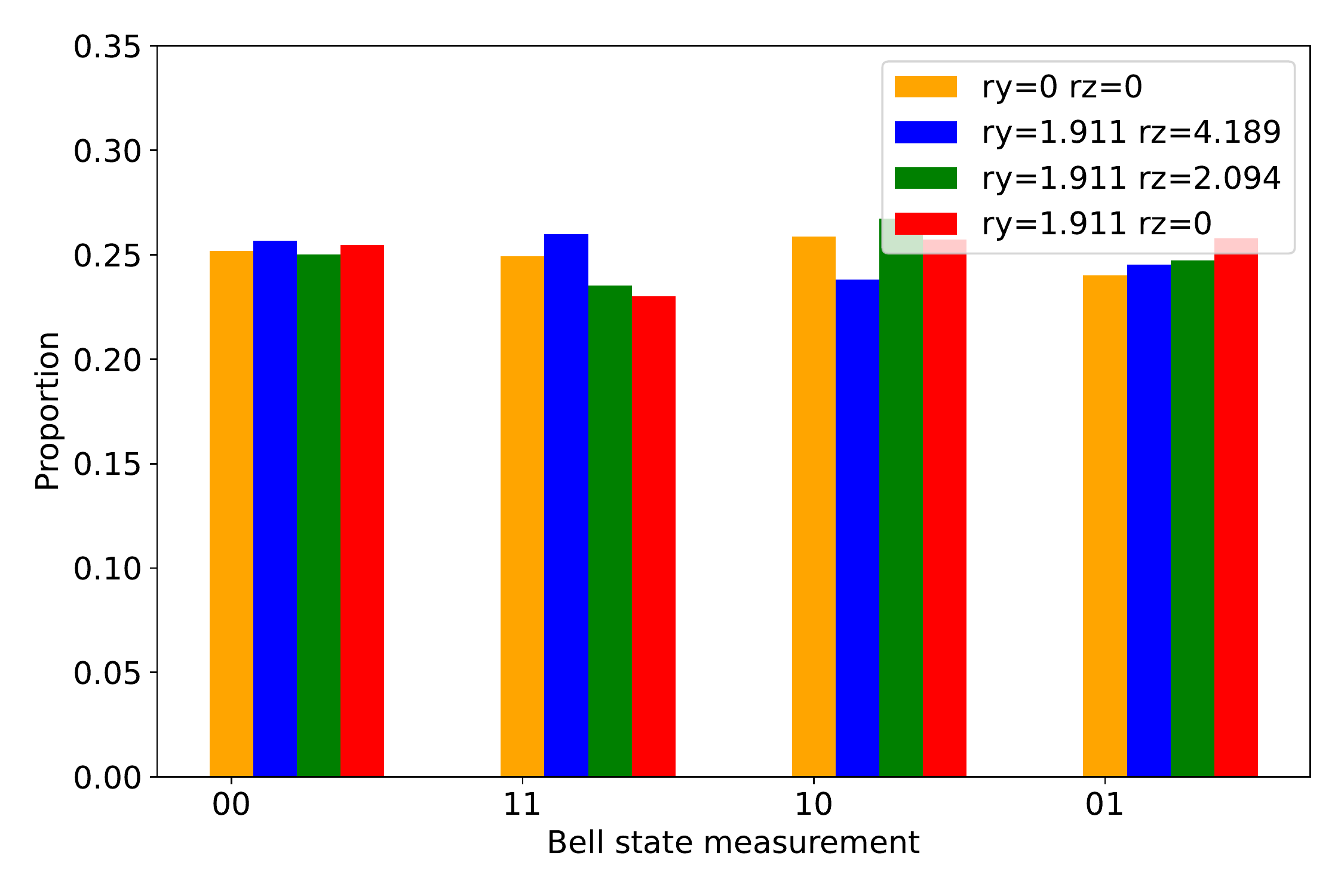}
    \caption{\textbf{Left:} Computed clone fidelities for each of the four cloned states as well as the ideal clone fidelity. Notice that the ideal clone fidelity (given by Eq. \ref{eq:theoretical-fidelity}) is the same for all clones since the telecloning is symmetric. \textbf{Right:} Mid-circuit Bell state measurement proportions. }
    \label{fig:result_barplots}
\end{figure*}

\begin{figure}[h!]
    \centering
    \includegraphics[width=0.45\textwidth]{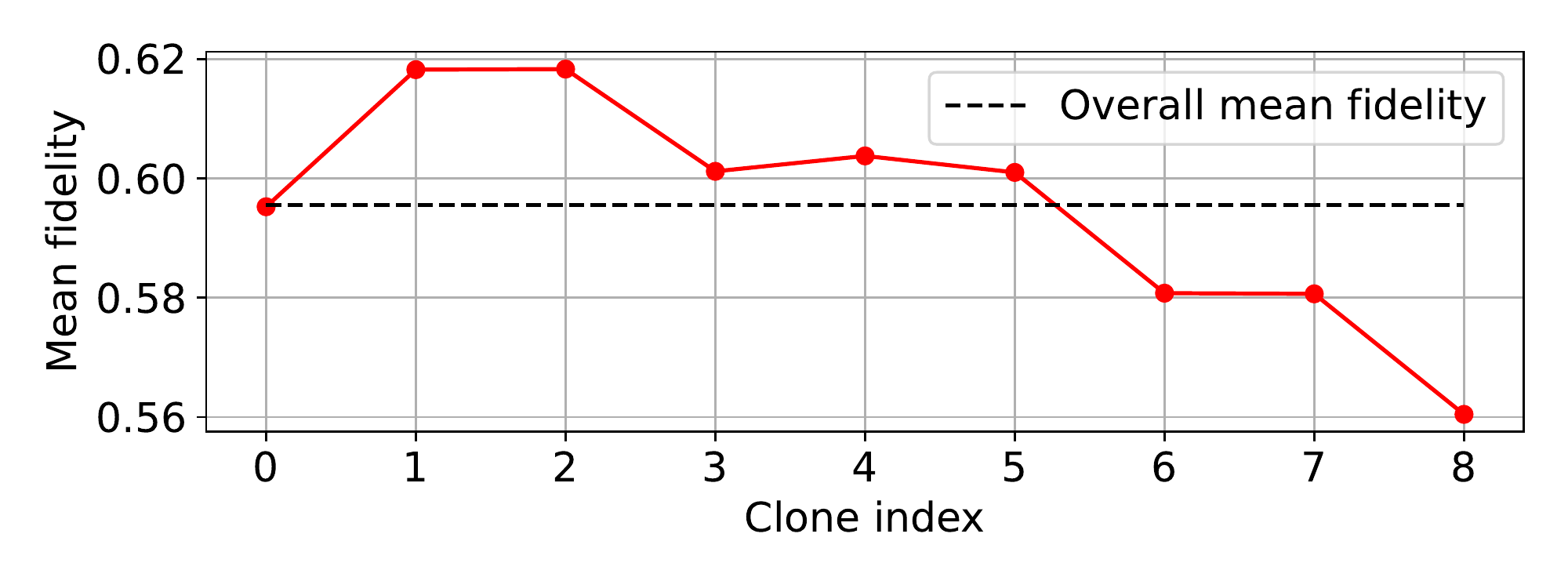}\\
    \vspace{-0.2 cm}
    \includegraphics[width=0.45\textwidth]{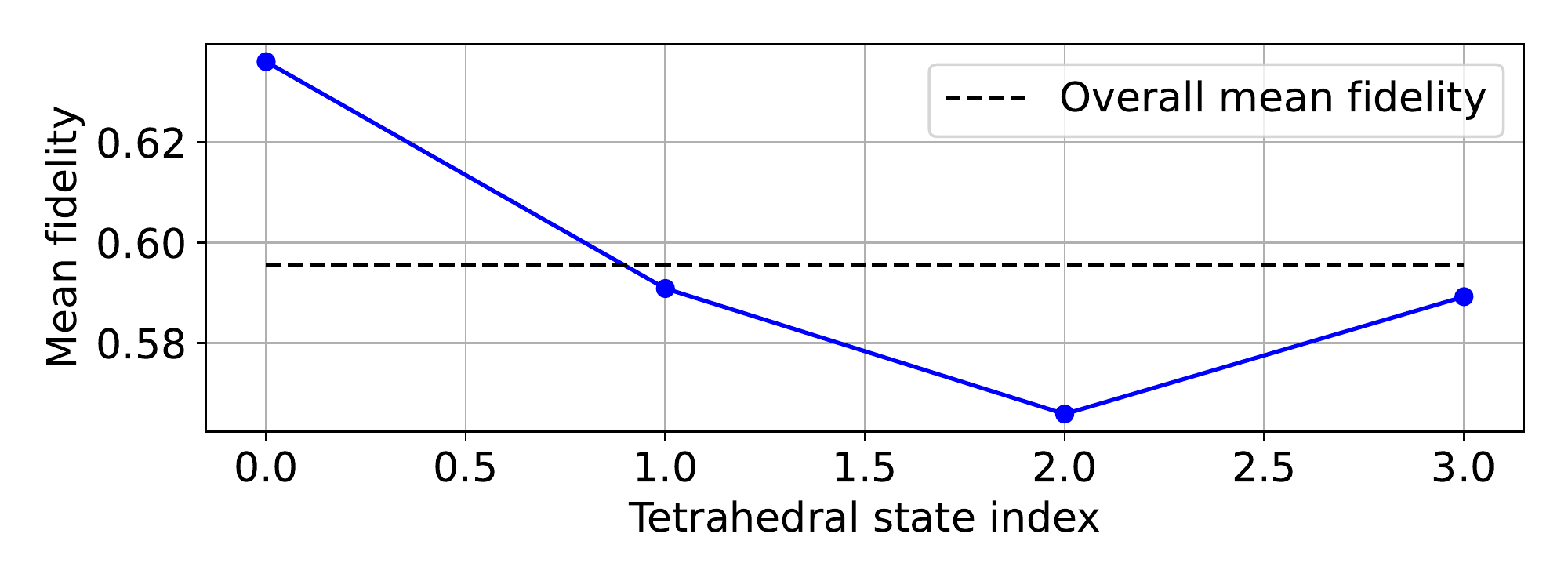}
    \caption{ Summary plot of Figure \ref{fig:result_barplots} for mean teleclone fidelity over the $9$ different clones and over the $4$ input message states. Horizontal dashes black lines indicate the overall clone fidelity. The tetrahedral state indices follow the same top-down order in the legend of the left hand plot in Figure \ref{fig:result_barplots}. Clone indices are consistent with the top-down order in Figures \ref{fig:main-circuit} and \ref{fig:M9_circuit}. }
    \label{fig:results_averages}
\end{figure}

At the time of executing these circuits on H1-1, the vendor provided typical single qubit error rate is $4 \cdot 10^{-5}$ and the typical two qubit error rate is $3 \cdot 10^{-3}$. The typical state preparation and measurement (SPAM) error rate is $3 \cdot 10^{-3}$.

Figure \ref{fig:M9_circuit} shows the circuit drawing for the $1 \rightarrow 9$ telecloning circuit. Currently the Quantinuum H1-1 device contains exactly $20$ qubits, and the $1 \rightarrow 9$ telecloning circuit requires $19$ qubits, therefore this $9$ clone construction is the largest telecloning circuit that can be executed on H1-1, which is the largest publicly available quantum computer with if-statement and mid-circuit measurement capabilities.

In order to quantify the experimental clone quality, we will use the metric of \emph{fidelity} \cite{jozsa1994fidelity}, defined as follows: 
\begin{equation}
    F(\rho_1, \rho_2) = Tr[ \sqrt{ \sqrt{\rho_1}\rho_2 \sqrt{\rho_1} } ]^2
    \label{eq:fidelity-general}
\end{equation}
As shown in Eq. \ref{eq:fidelity-general}, we can compute the fidelity given the pure quantum state density matrix and the experimental computed density matrix. In order to compute the density matrix of the experimental results, we utilize \emph{parallel single qubit state fidelity} using the Pauli basis and a least-squares maximum likelihood estimation (MLE) fitter in a slightly modified version of Qiskit Ignis \cite{Qiskit, PhysRevLett.108.070502}. The key question that our experiments answer is how individual clone fidelity compares to the theoretical fidelity of the telecloning process.

In order to implement \emph{parallel single qubit state fidelity}, we prepare three versions of the each telecloning circuit - one version for each of the Pauli basis (X, Y, and Z). Since we only need to quantify the individual clone fidelity, we can simply apply a Pauli basis rotation to each of the clone qubits and then repeat this measurement for all three Pauli basis. This also allows us to parallelize the state tomography computation across all of the clone qubits since we are only applying single qubit state tomography. This approach saves a considerable amount of computation time when compared to standard full state tomography of 
 with  $3^n$ runs to test all possible Pauli bases. Here $n$ is the system size that the state tomography would be applied to, which for $M=9$ is $19,683$. Parallel single qubit state tomography is extraordinarily convenient for quantifying the clone quality of large clone numbers as the number of separate basis state executions we need is always exactly $3$ for Pauli basis state tomography. 

When executing the circuits on H1-1, $500$ shots are used for each circuit. All other parameter settings are left to default values. In particular,  server side optimization is turned on which means that the device can compile and optimize the circuit before executing it. The telecloning circuits (i.e., Figure \ref{fig:M9_circuit}) were constructed using Qiskit \cite{Qiskit}, then exported to OpenQASM \cite{https://doi.org/10.48550/arxiv.1707.03429} and then submitted to the Quantinuum H1-1 device. Figure \ref{fig:M9_circuit} shows specifically which gates are used to define the telecloning circuit, but these gates are not the native gateset of the H1-1 device and therefore part of the server side compilation must adapt the gateset. The native gates for the H1-1 are $R_z$, $U_{1q}$, and $ZZ$. 
\begin{figure*}[h!]
    \centering
        \rotatebox{90}{\smash{\small message $q_m$}}\ %
        \includegraphics[width=0.10\textwidth]{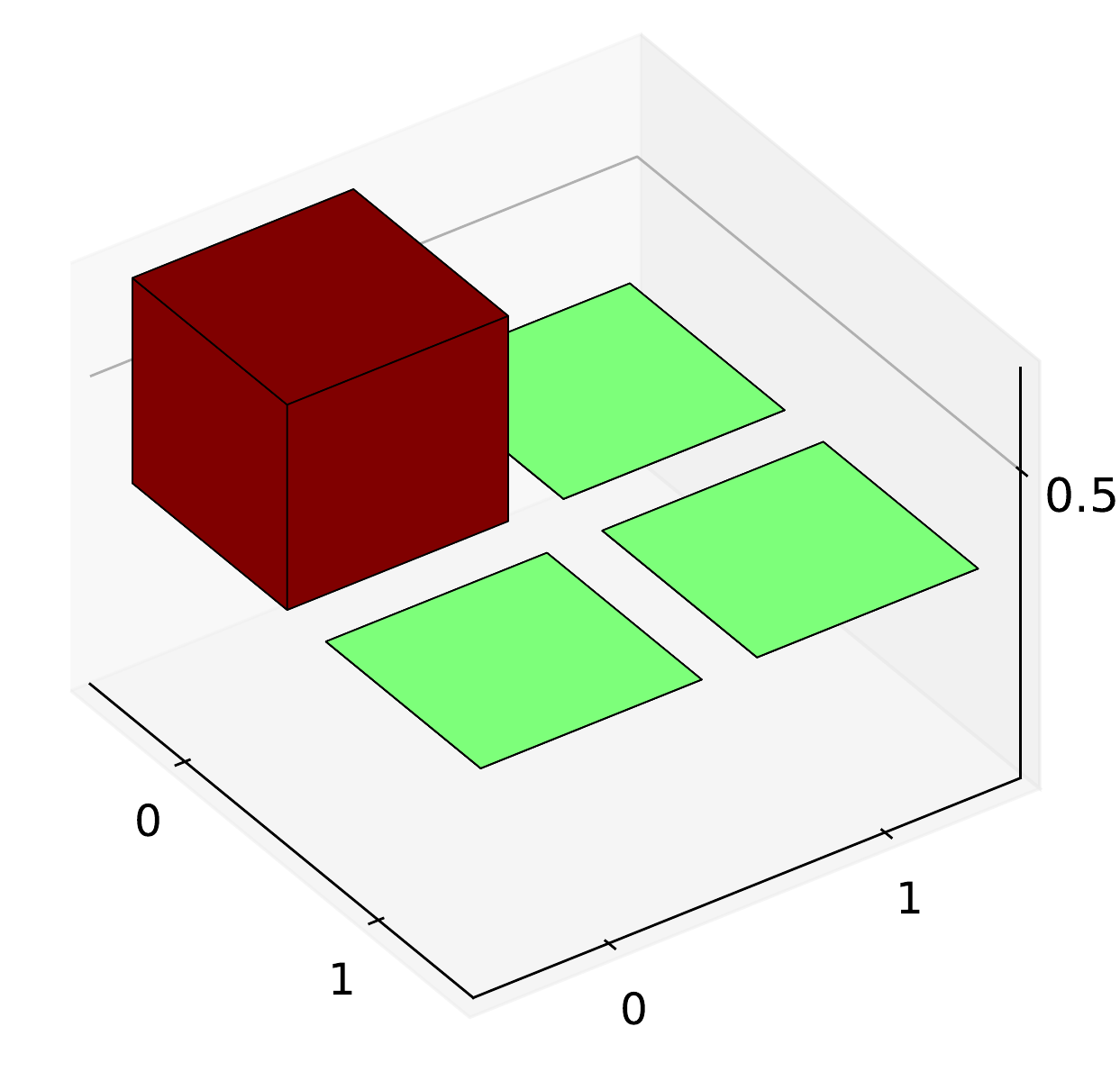}%
        \includegraphics[width=0.10\textwidth]{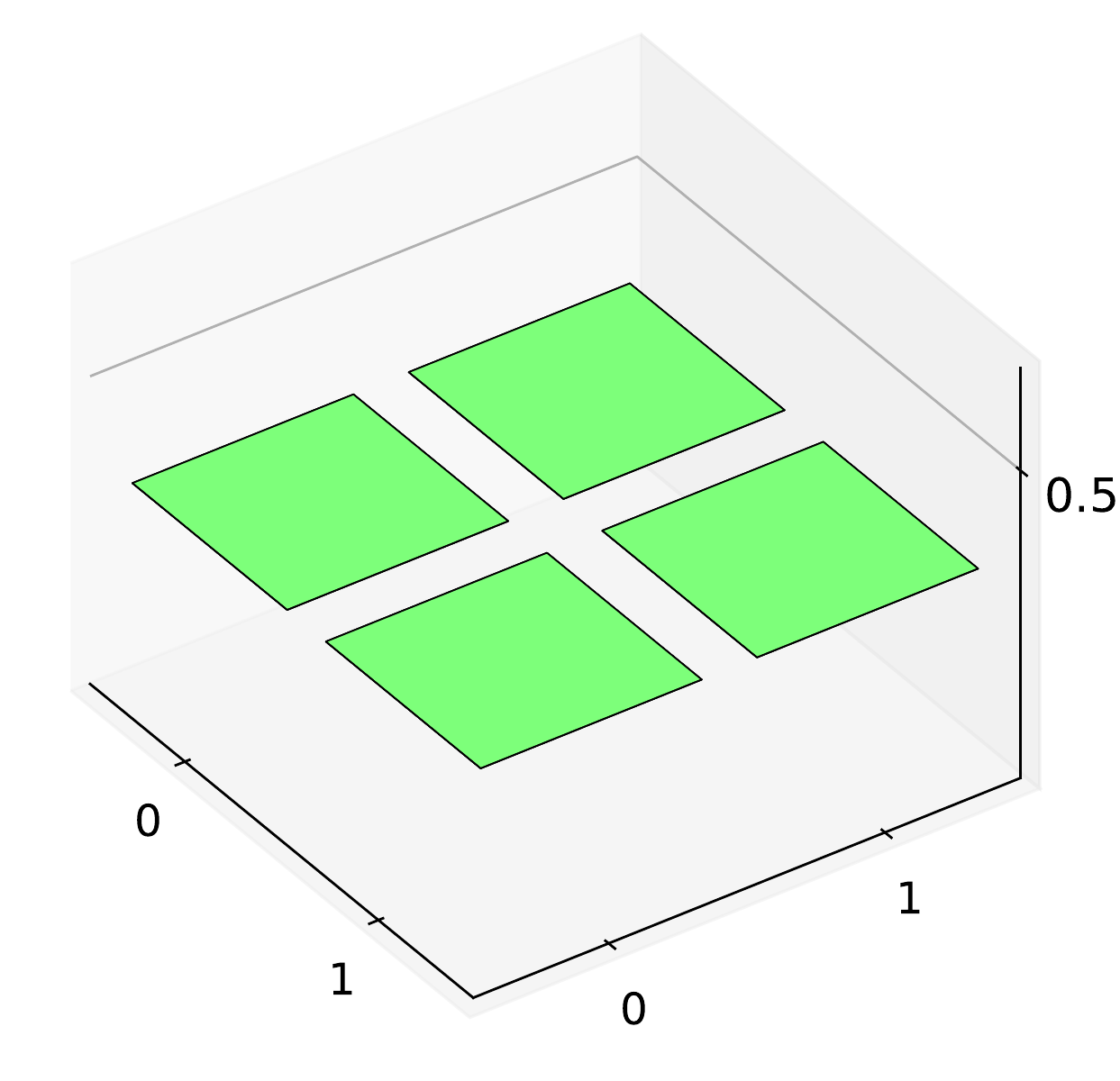}%
        \hfill%
        \includegraphics[width=0.10\textwidth]{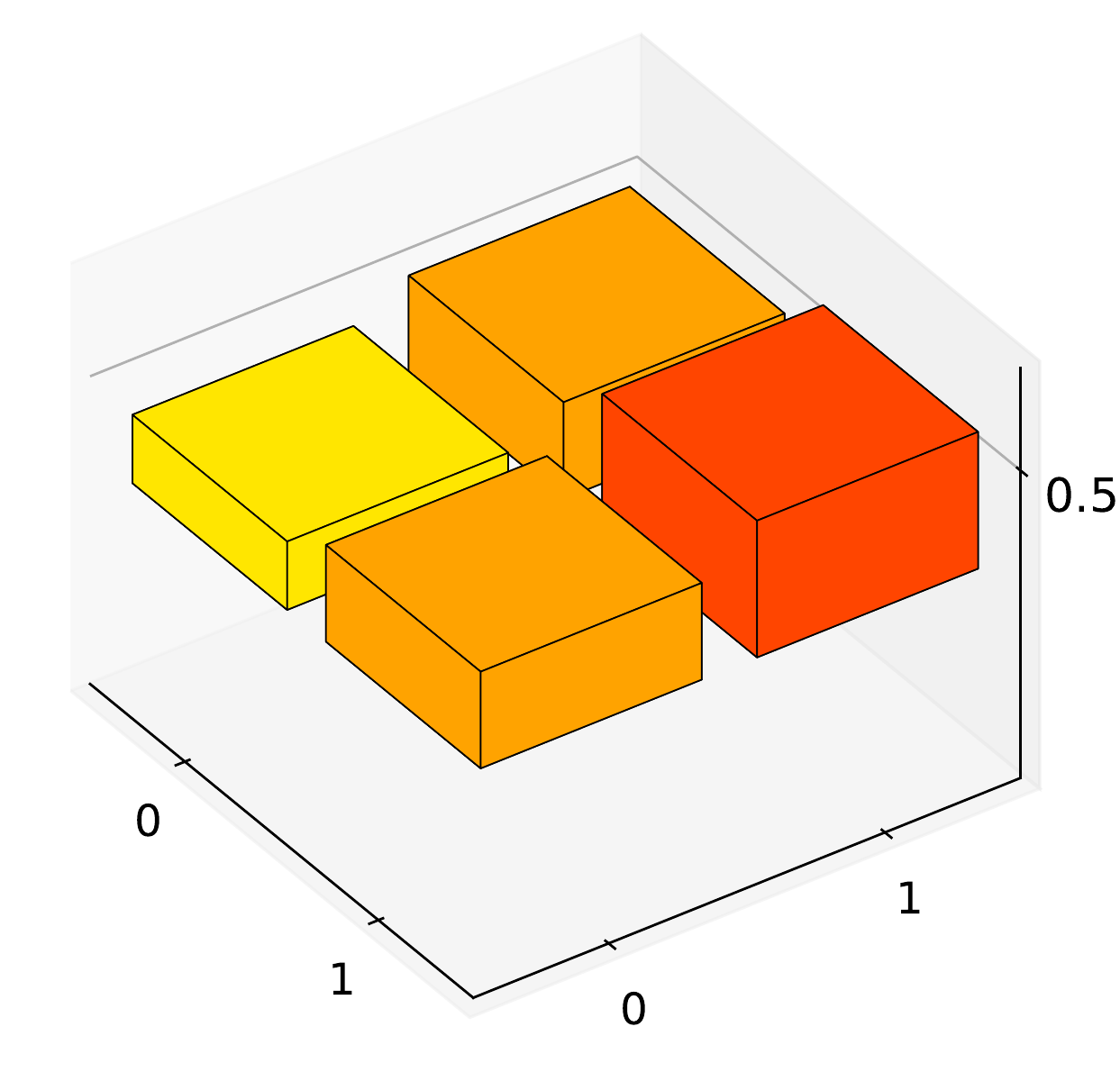}%
        \includegraphics[width=0.10\textwidth]{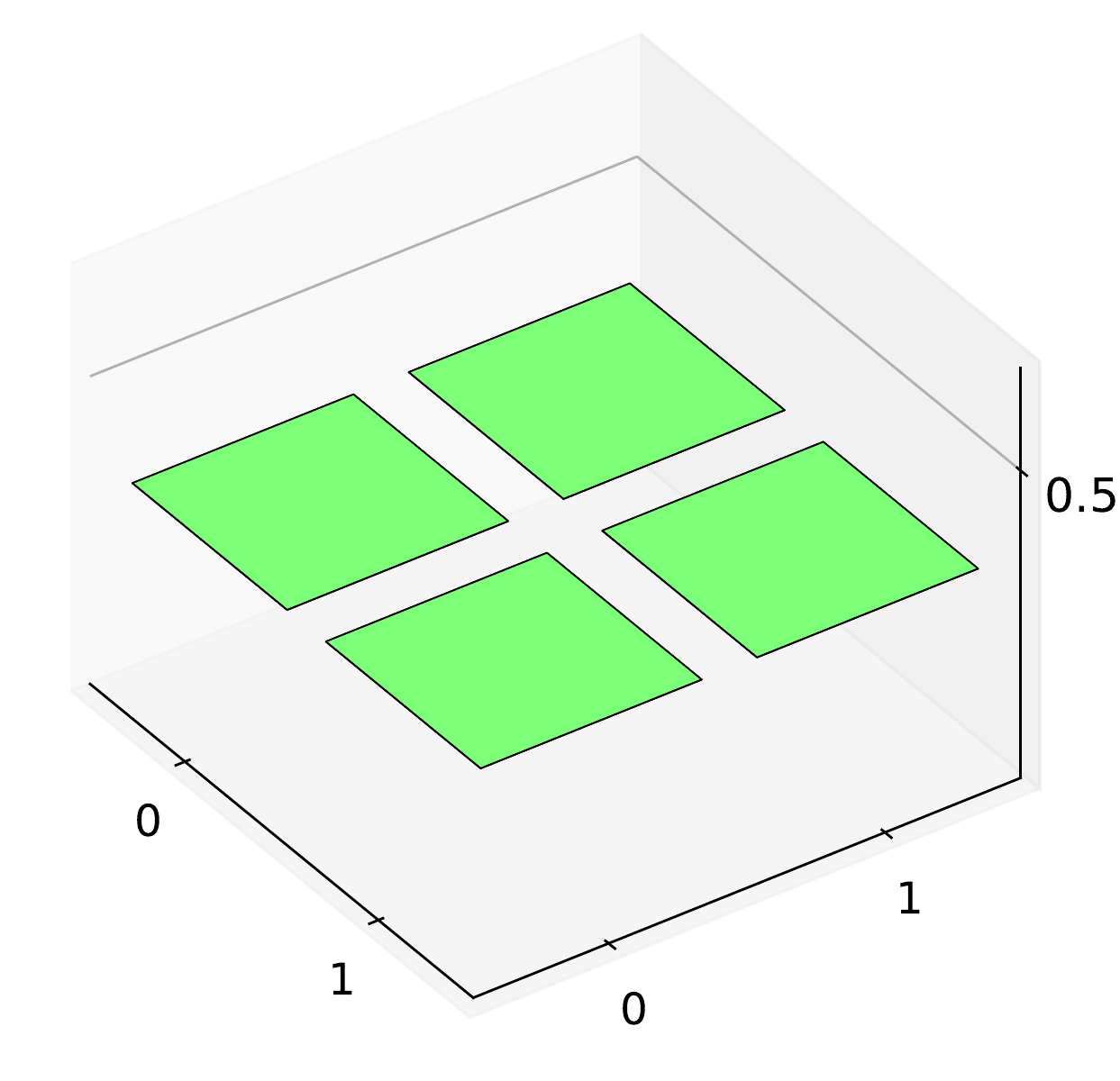}%
        \hfill%
        \includegraphics[width=0.10\textwidth]{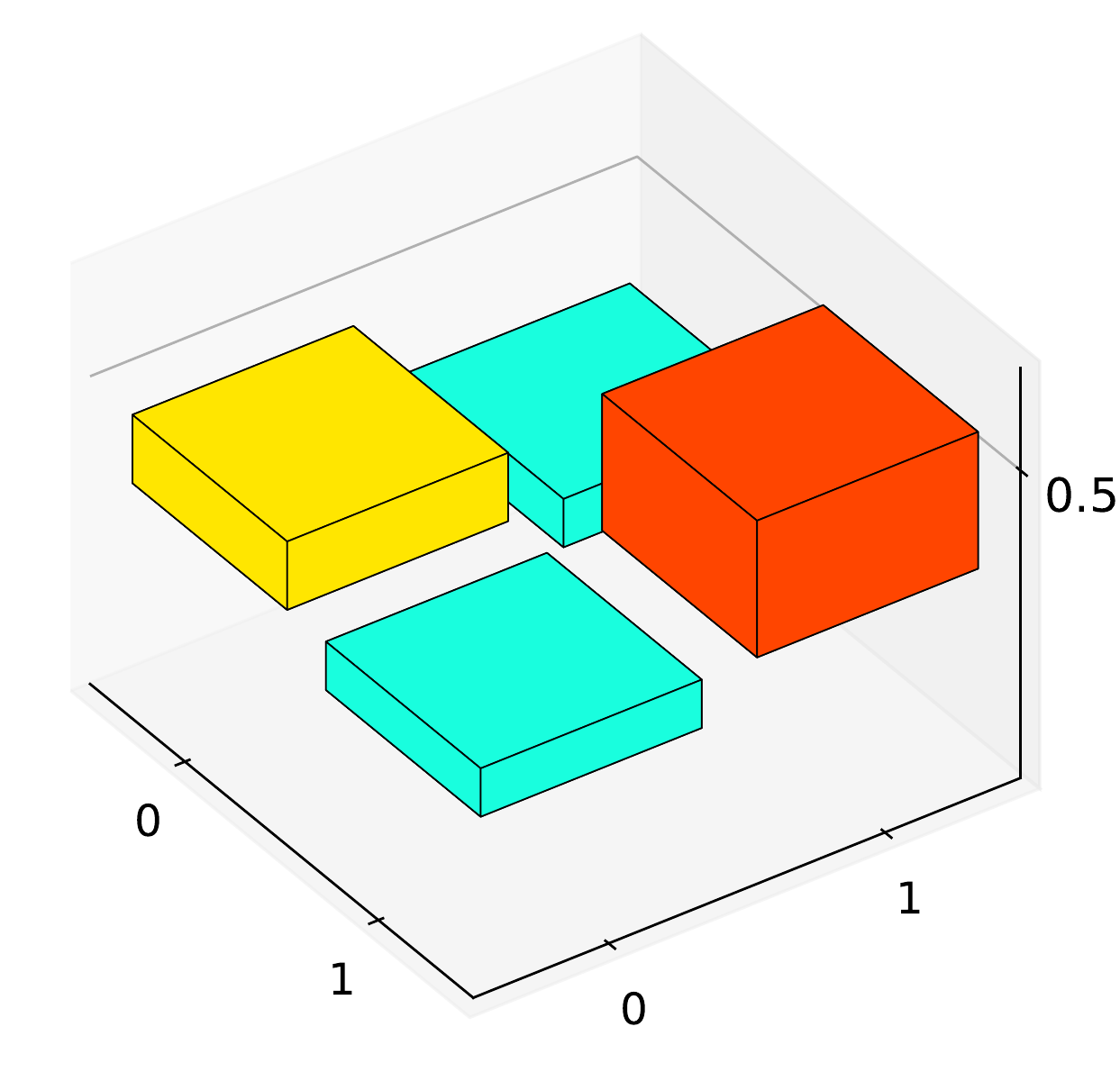}%
        \includegraphics[width=0.10\textwidth]{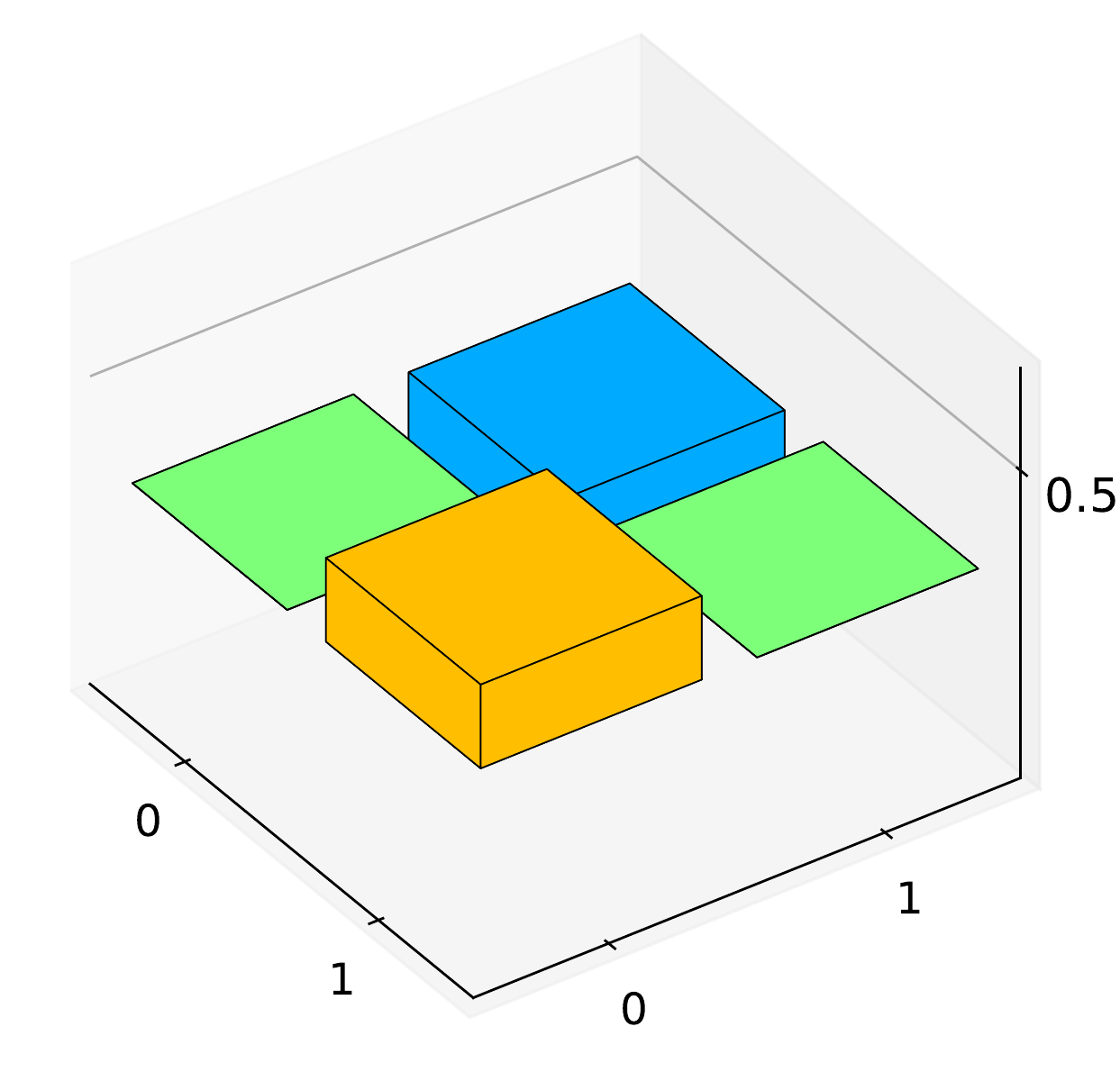}%
        \hfill%
        \includegraphics[width=0.10\textwidth]{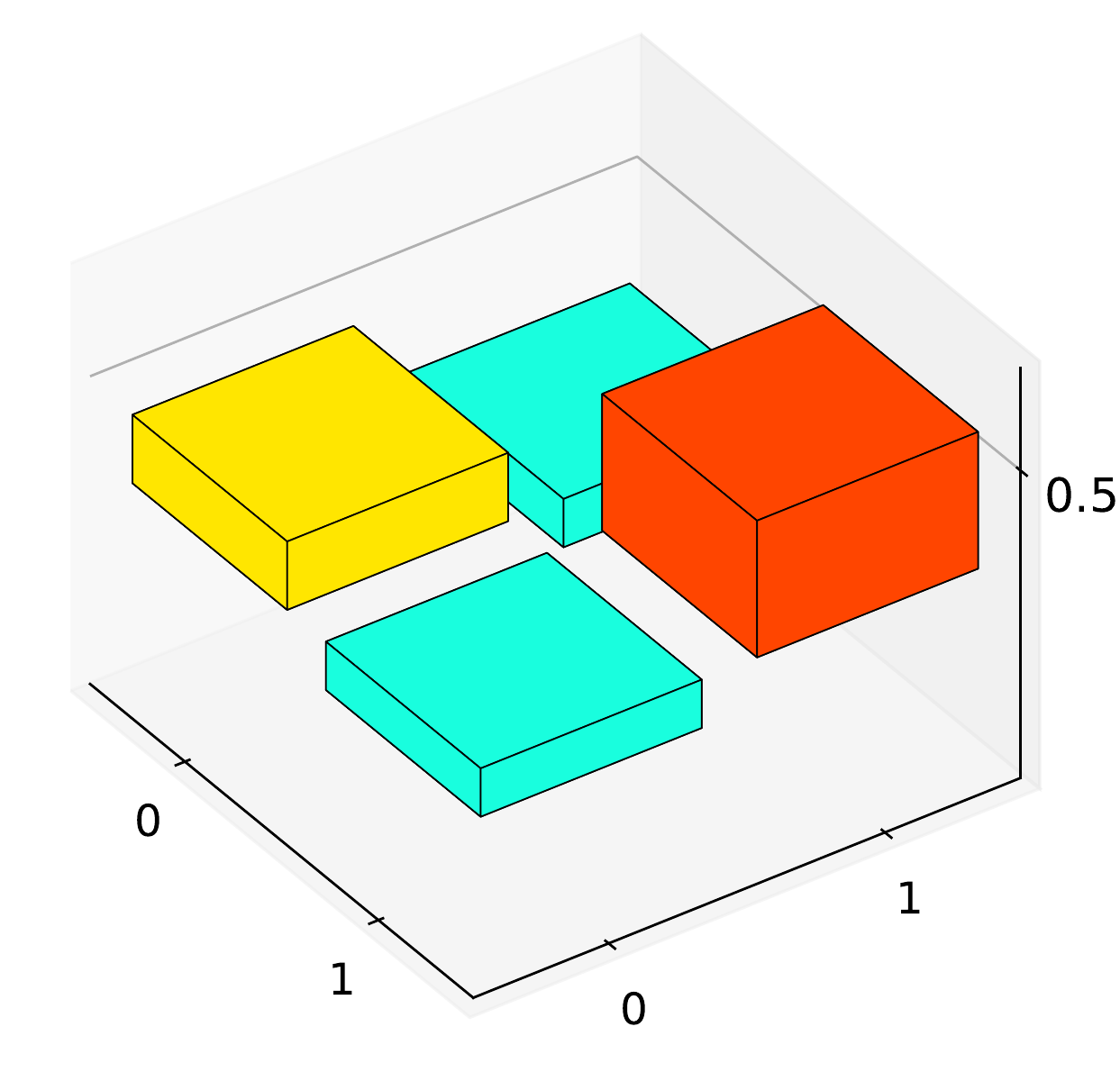}%
        \includegraphics[width=0.10\textwidth]{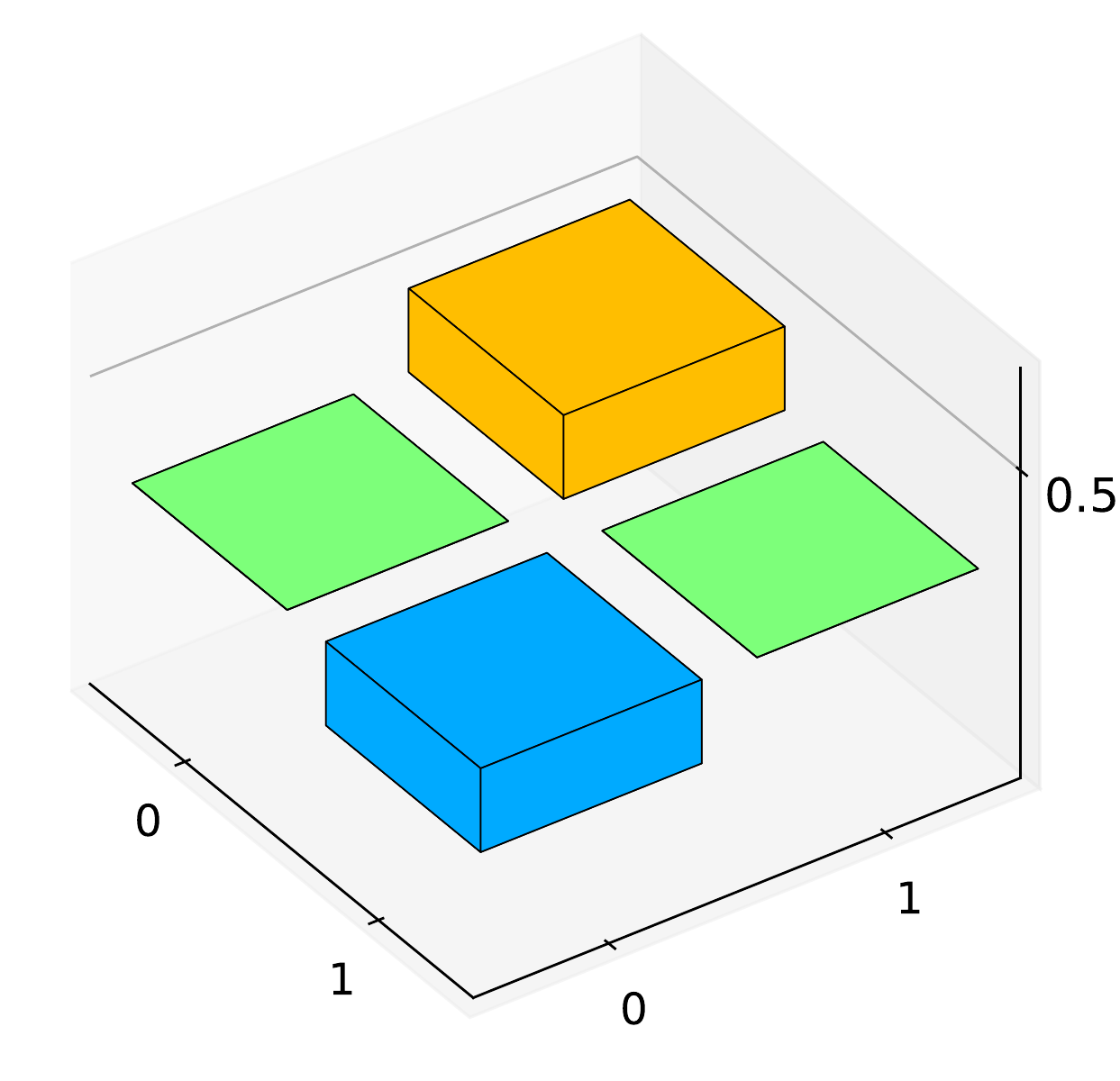}\\[1ex]%
        \rotatebox{90}{\smash{\small\ ideal clone}}\ %
        \includegraphics[width=0.10\textwidth]{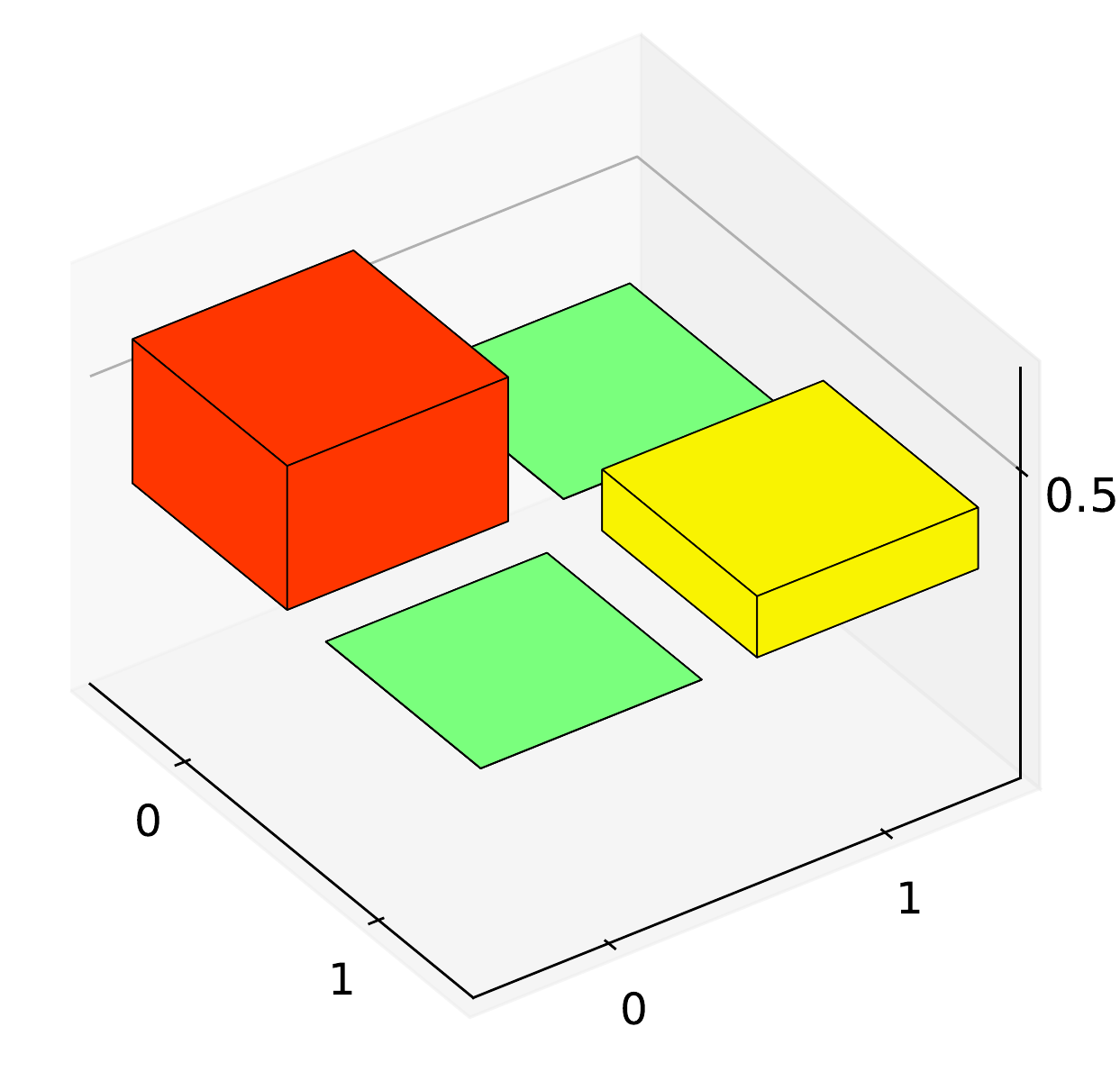}%
        \includegraphics[width=0.10\textwidth]{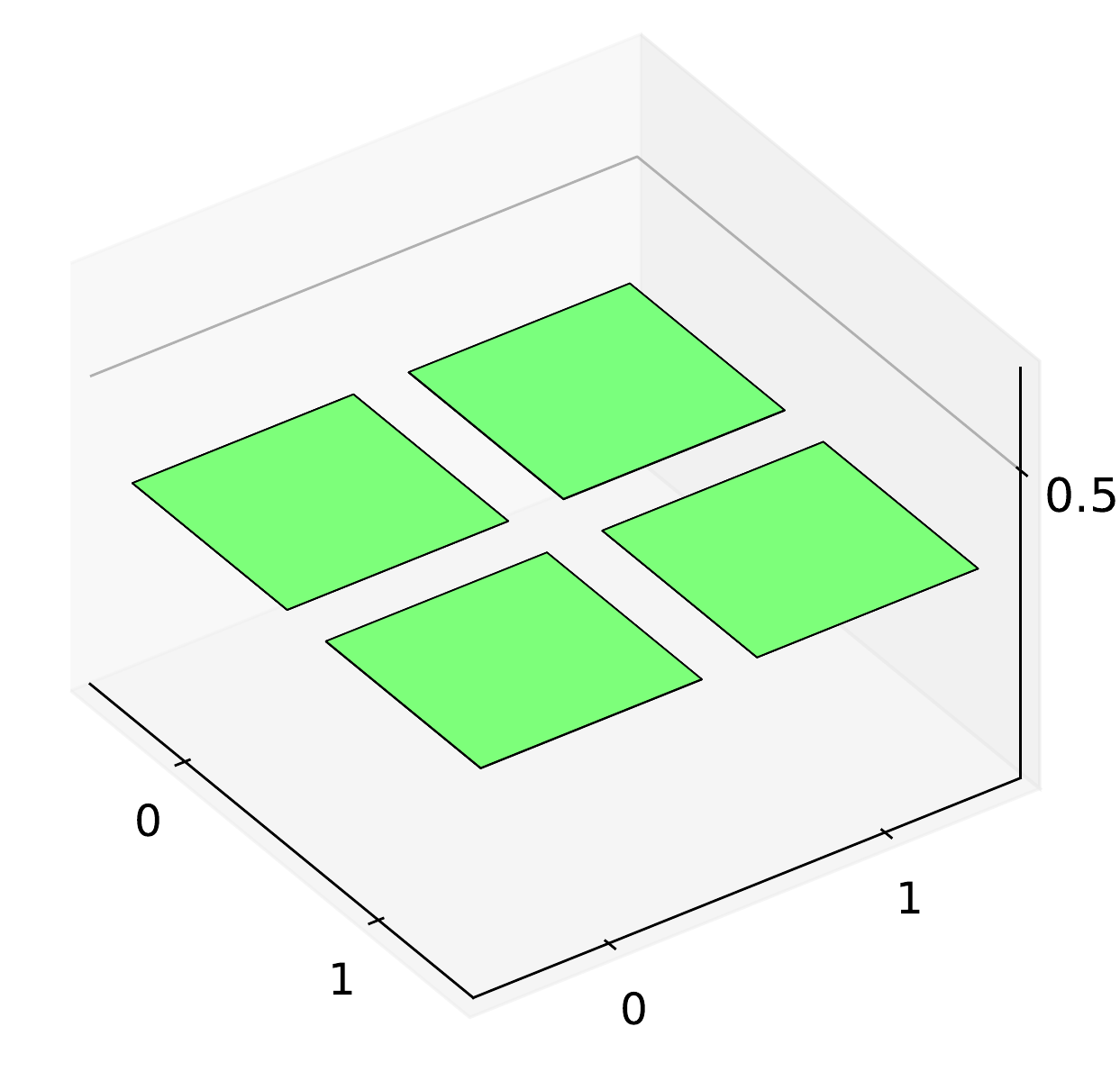}%
        \hfill%
        \includegraphics[width=0.10\textwidth]{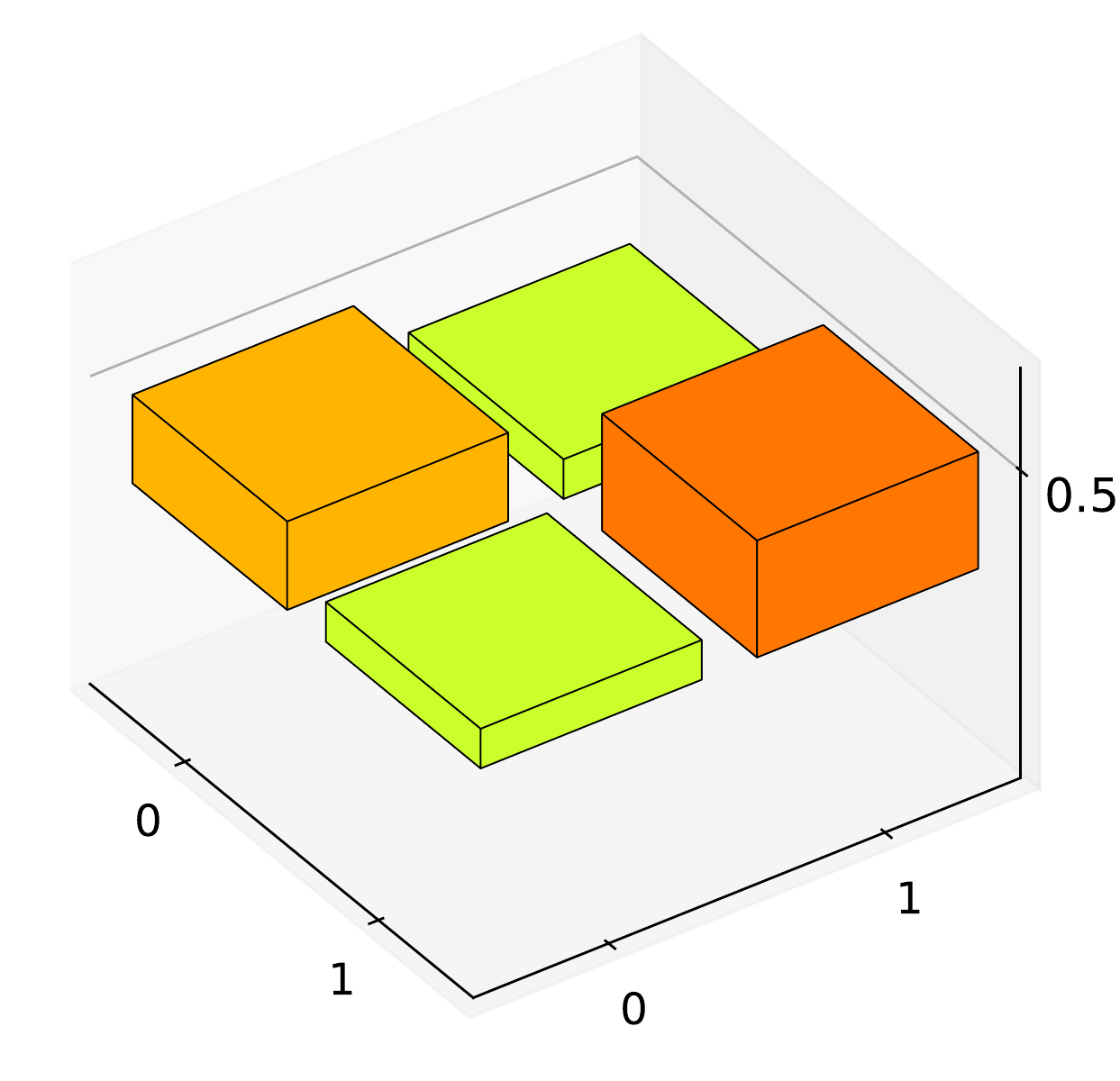}%
        \includegraphics[width=0.10\textwidth]{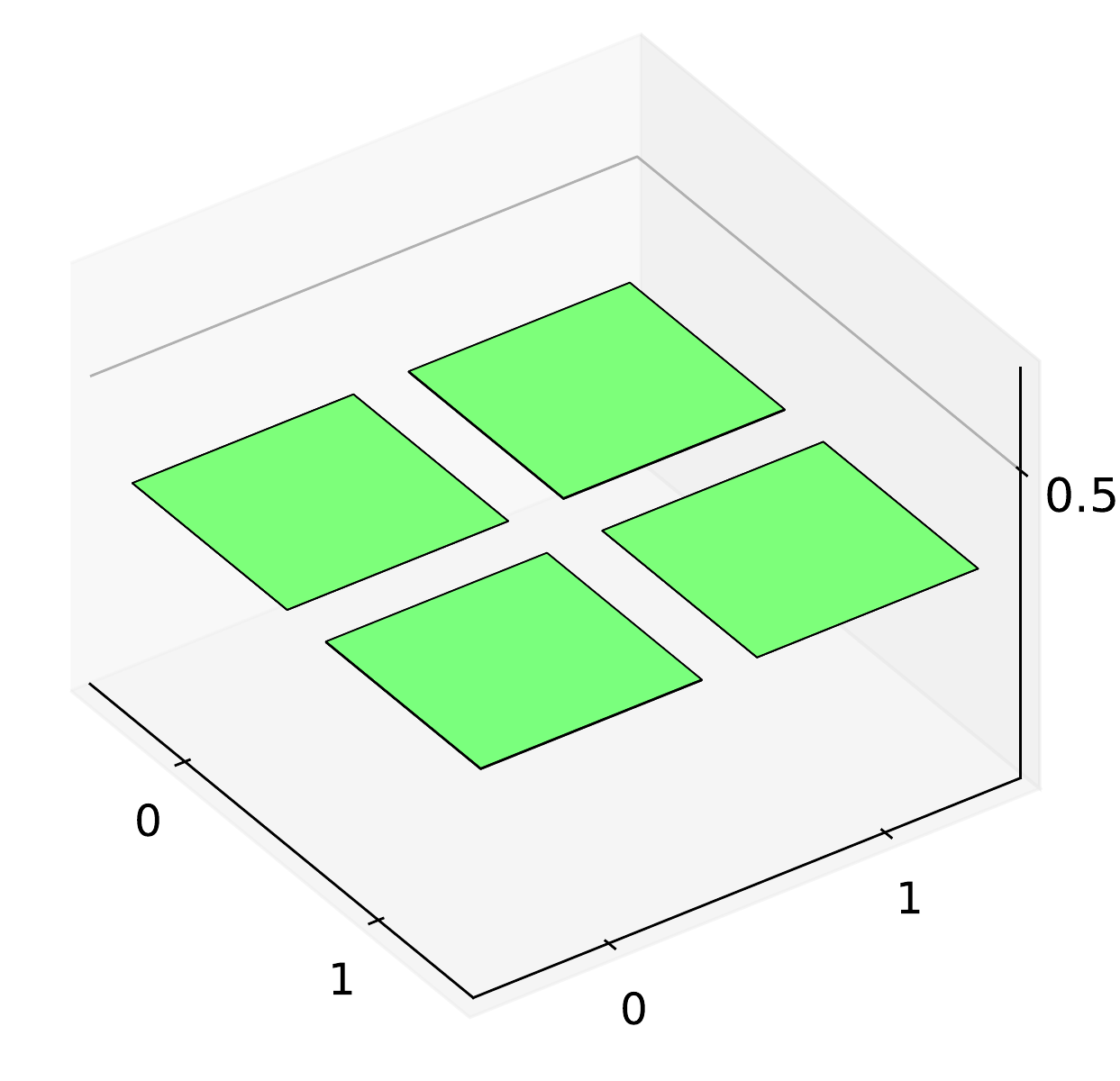}%
        \hfill%
        \includegraphics[width=0.10\textwidth]{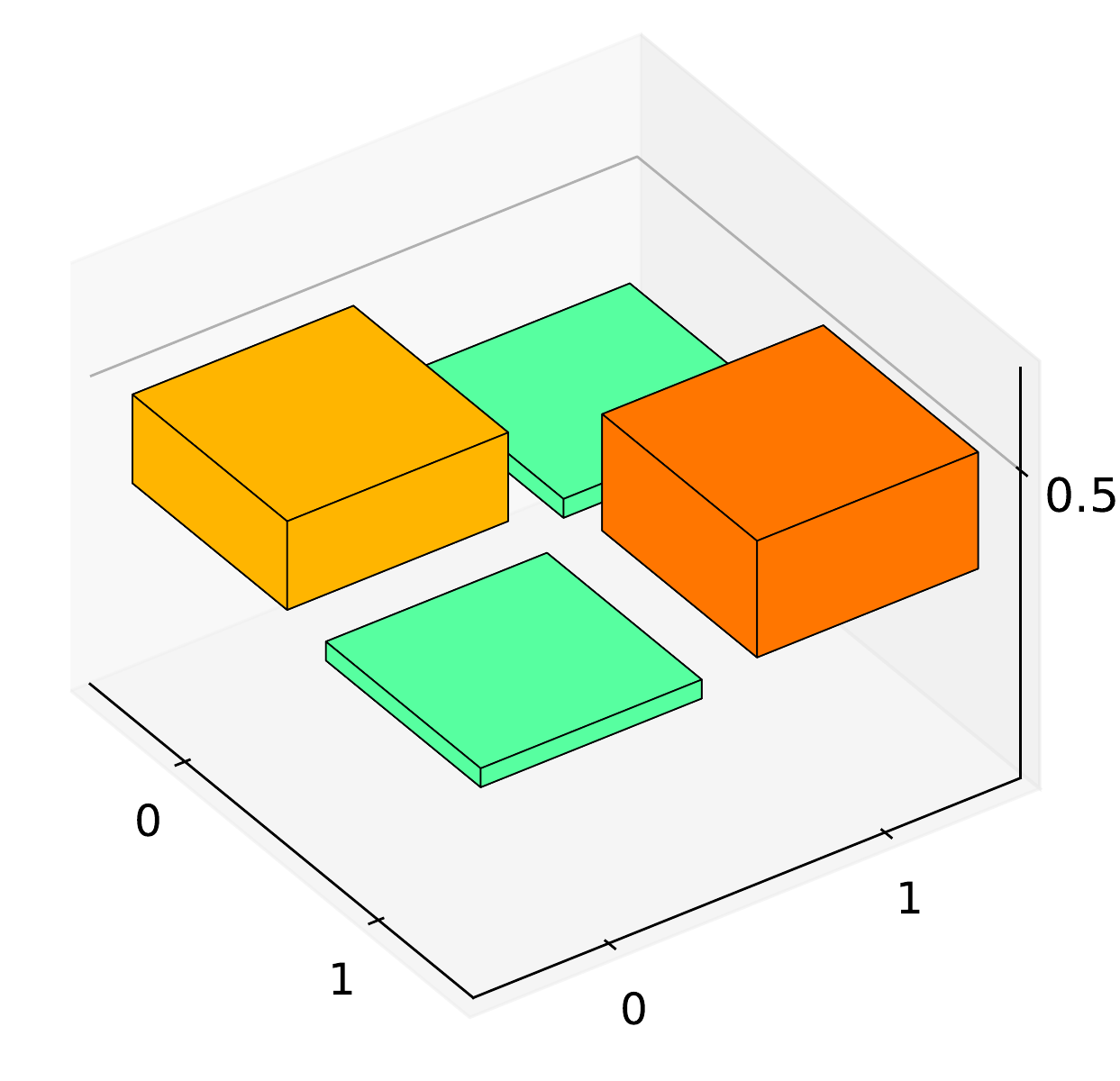}%
        \includegraphics[width=0.10\textwidth]{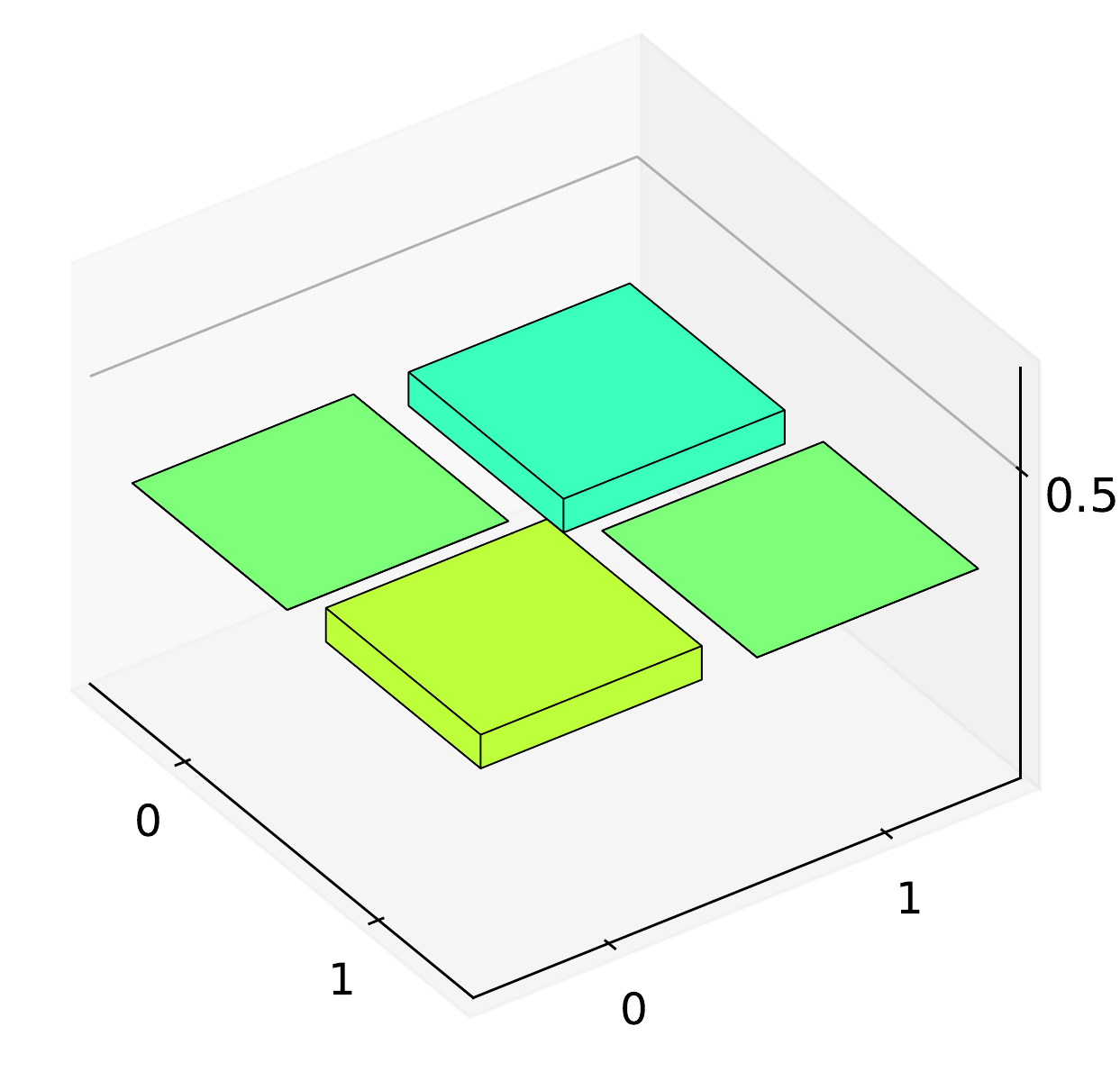}%
        \hfill%
        \includegraphics[width=0.10\textwidth]{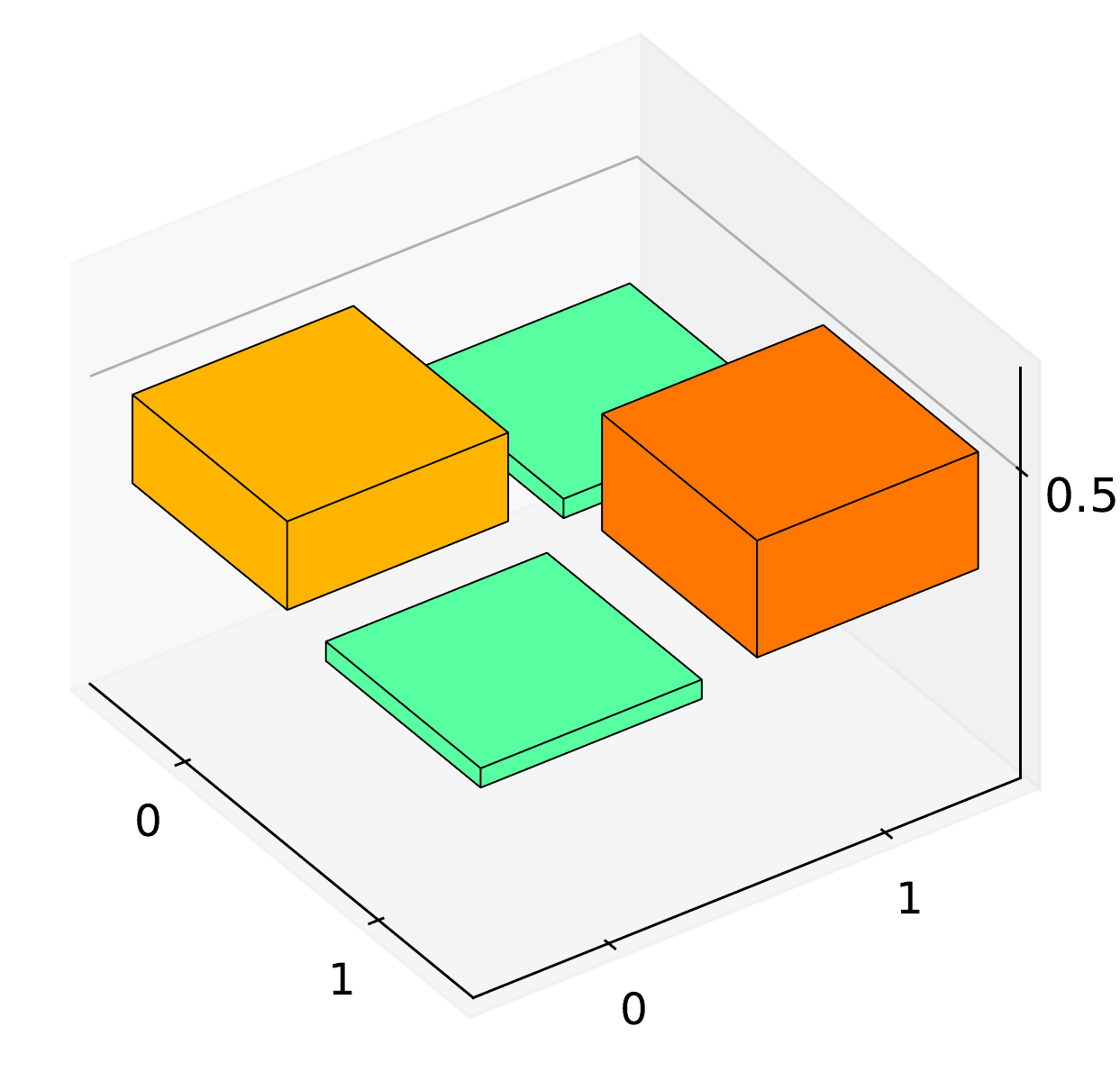}%
        \includegraphics[width=0.10\textwidth]{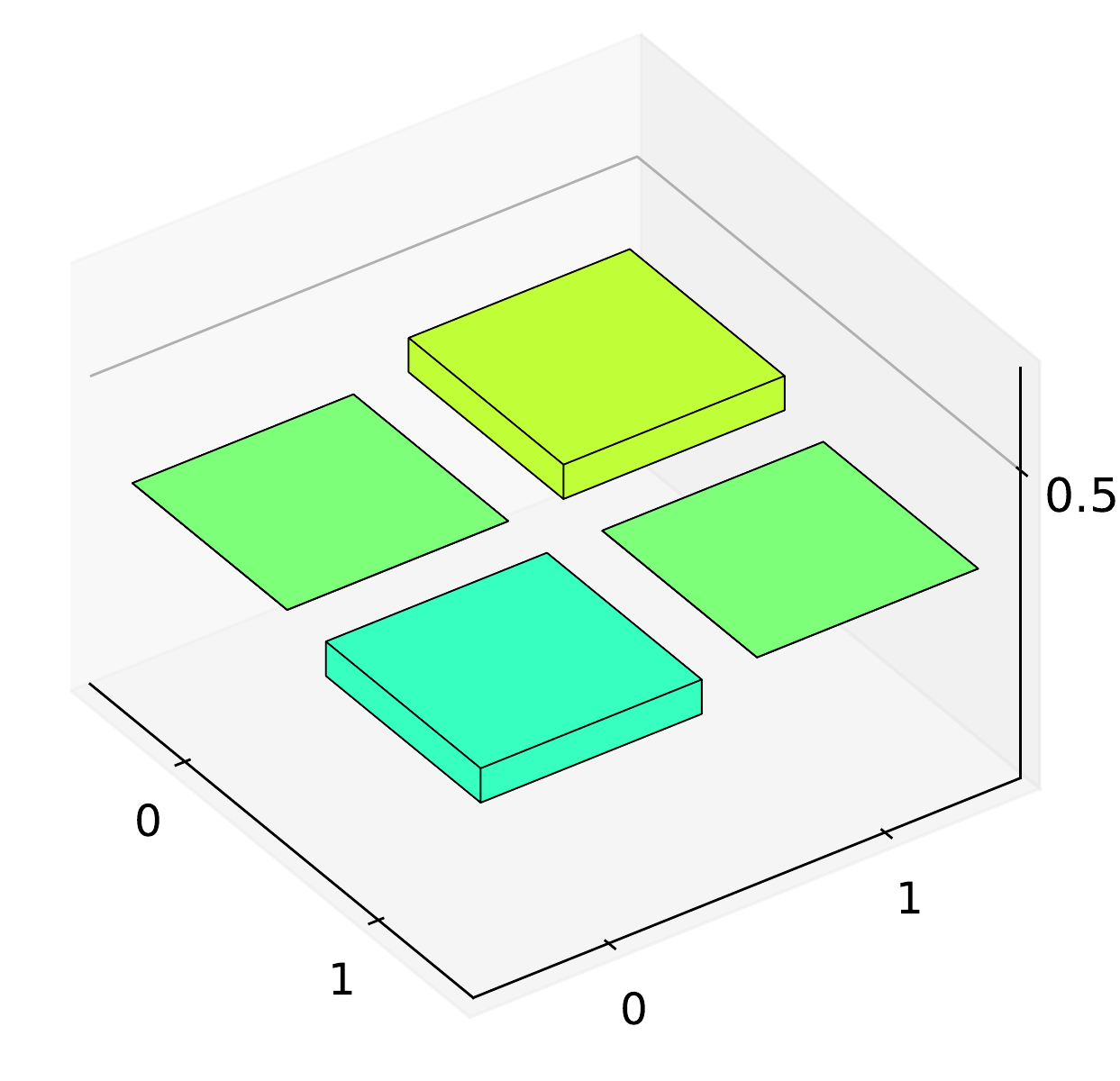}\\[1ex]%
        \rotatebox{90}{\smash{\small\ \ clone $C_0$}}\ %
        \includegraphics[width=0.10\textwidth]{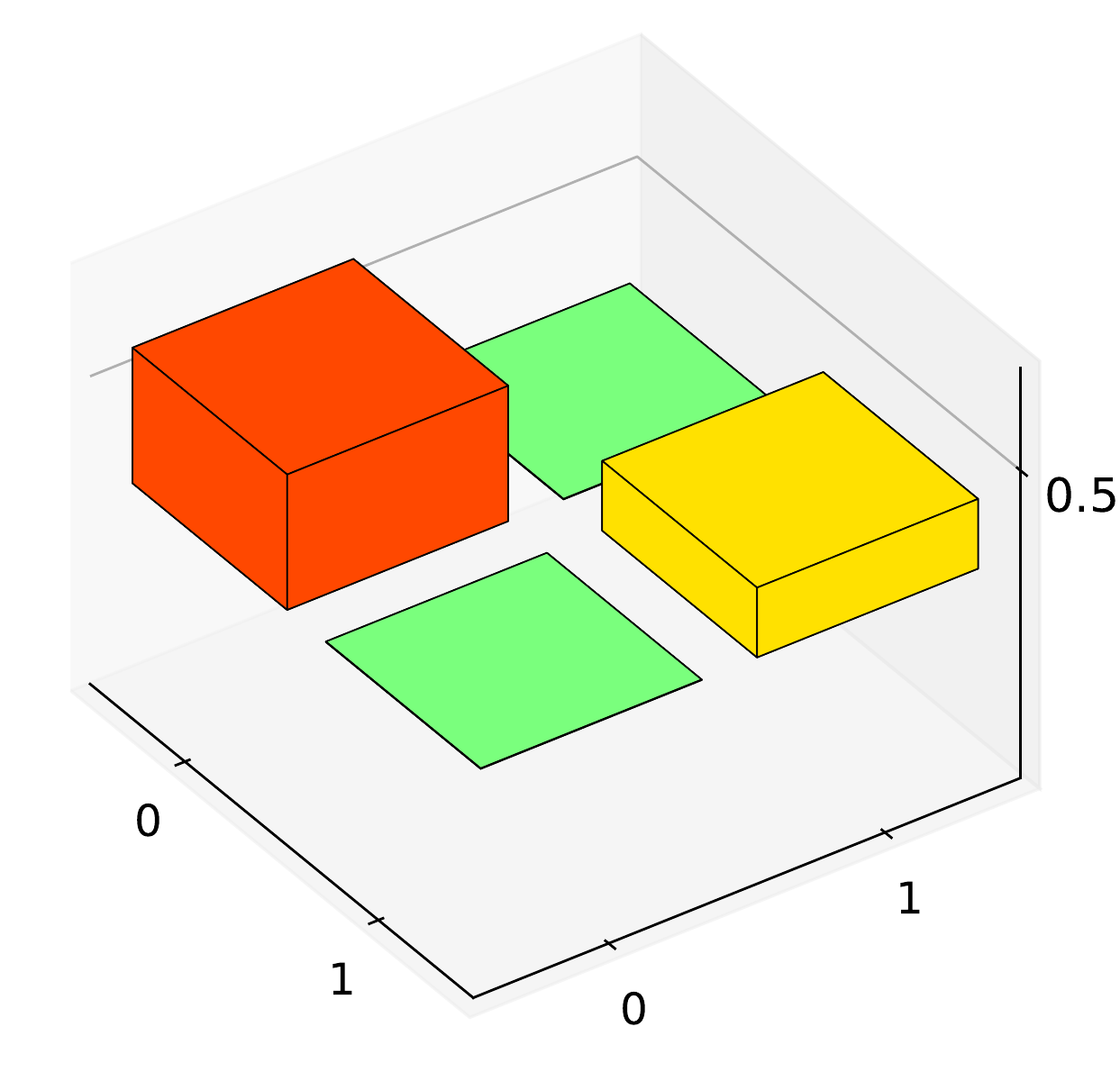}%
        \includegraphics[width=0.10\textwidth]{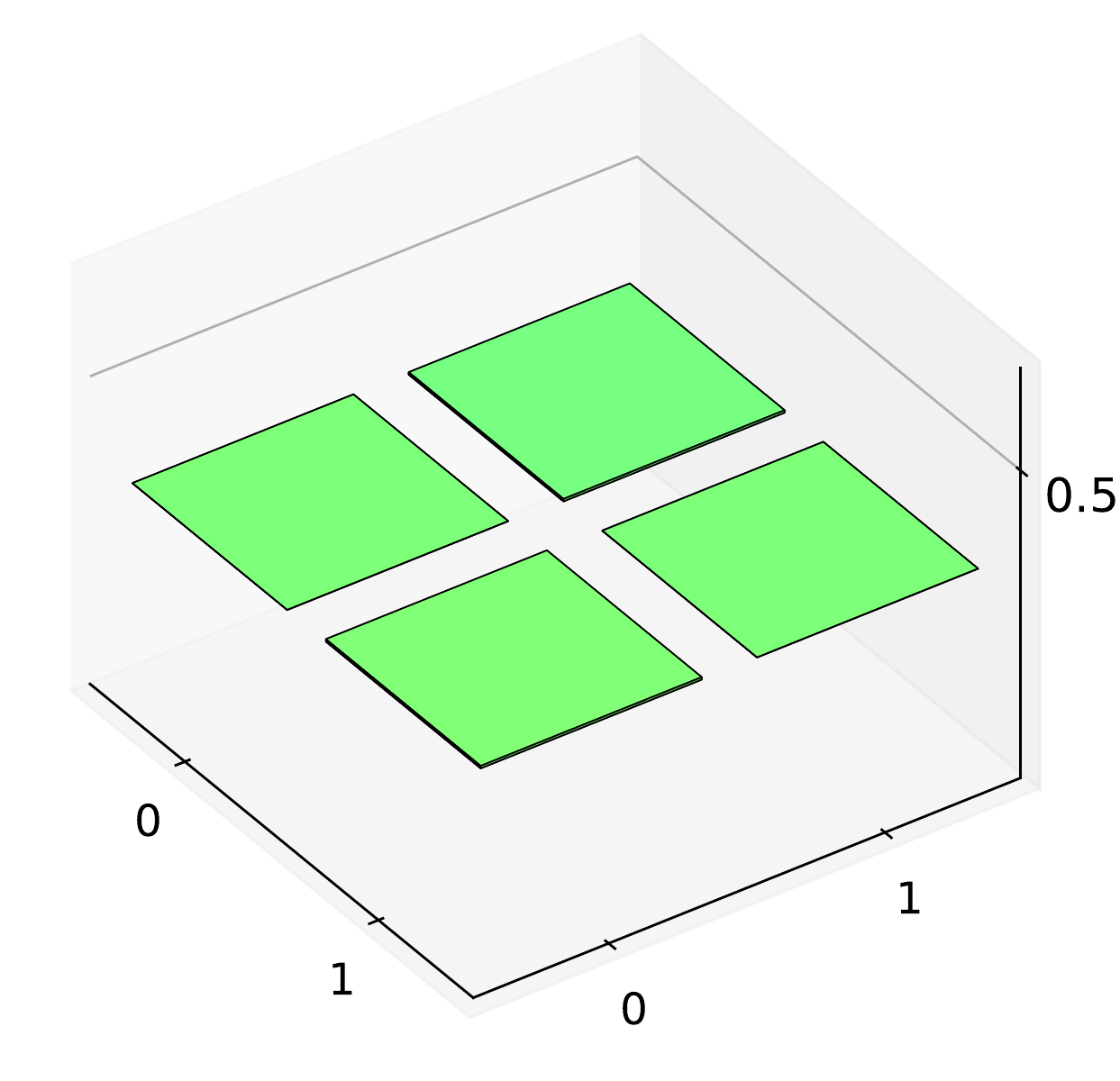}%
        \hfill%
        \includegraphics[width=0.10\textwidth]{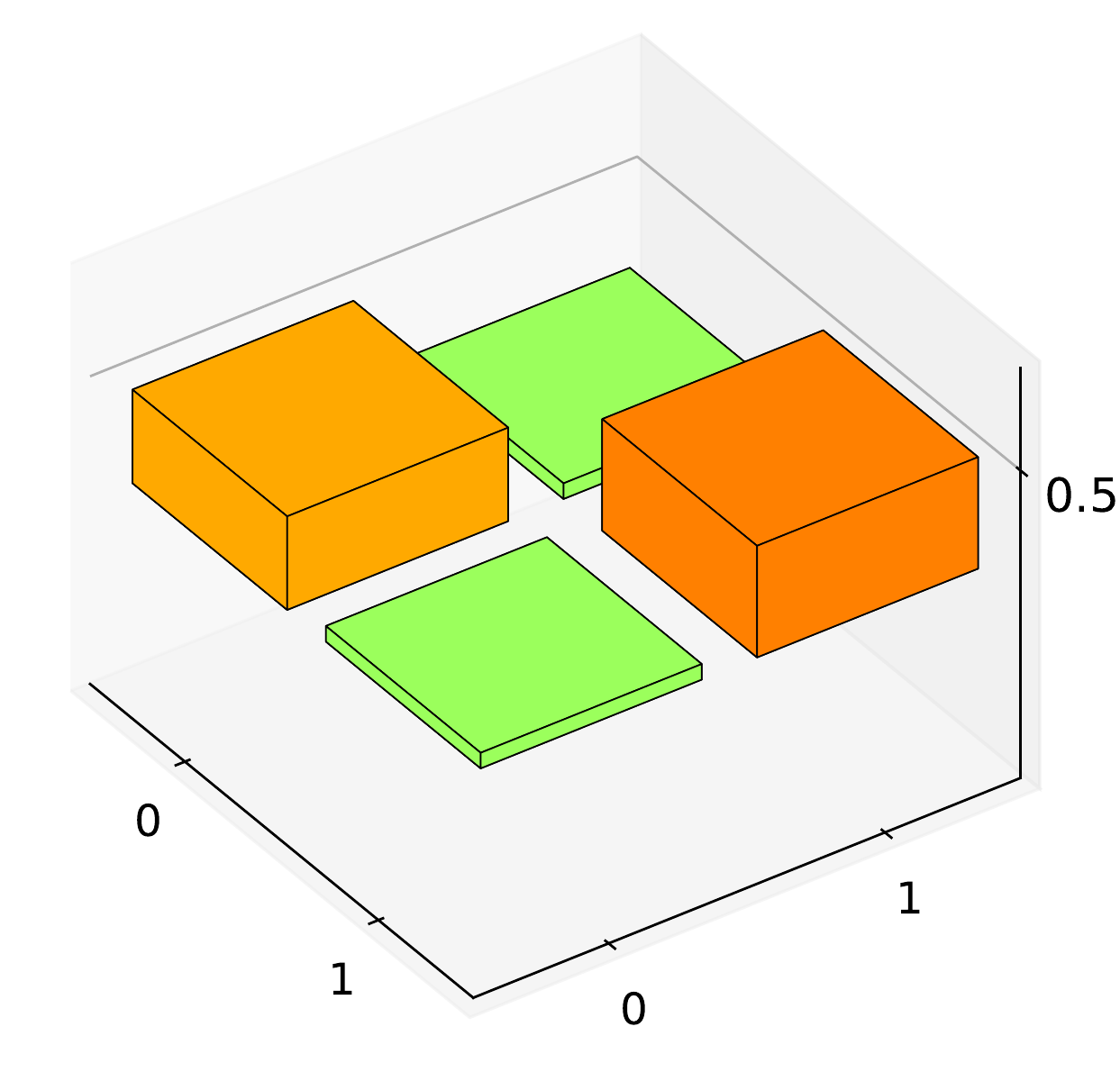}%
        \includegraphics[width=0.10\textwidth]{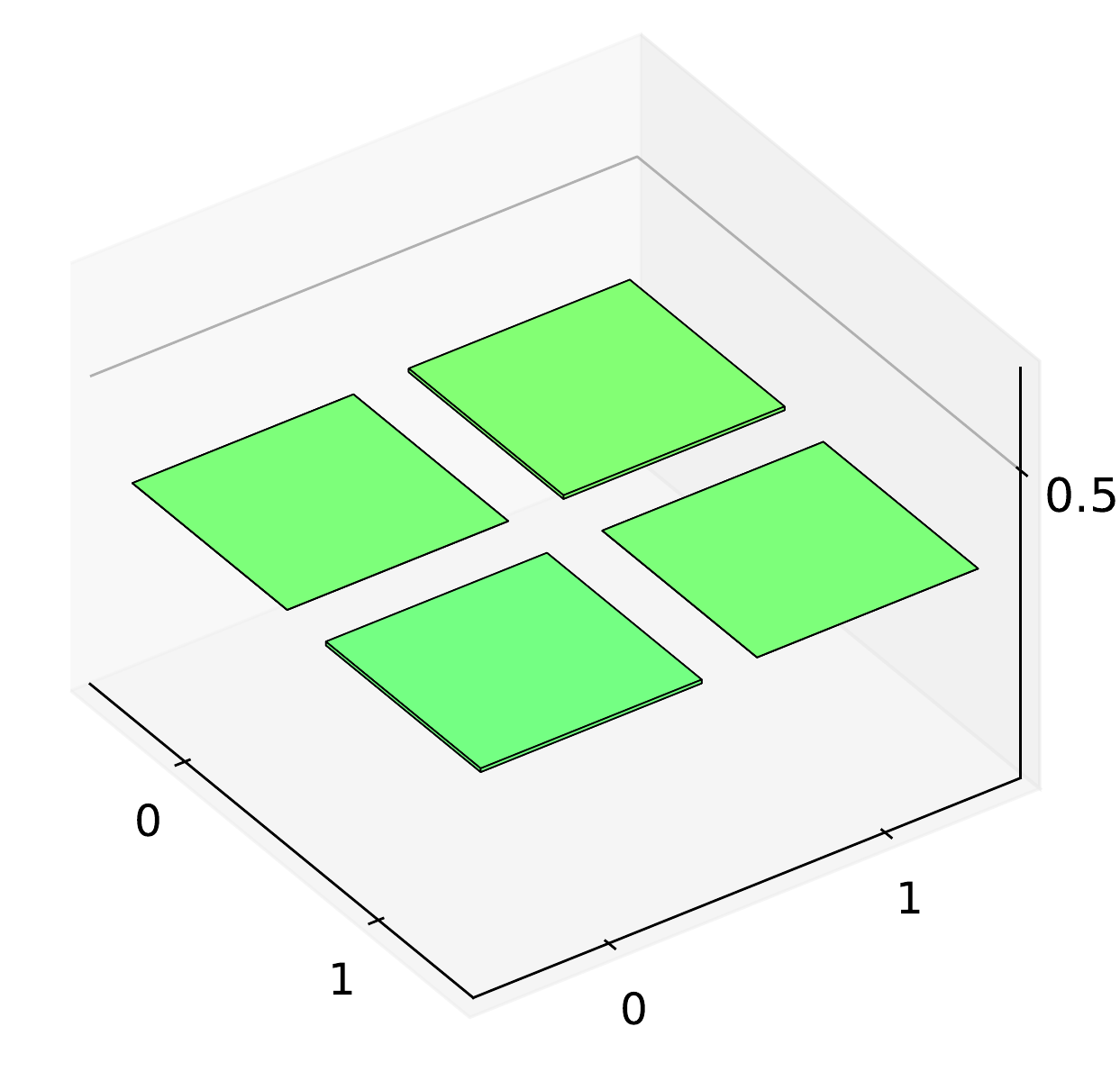}%
        \hfill%
        \includegraphics[width=0.10\textwidth]{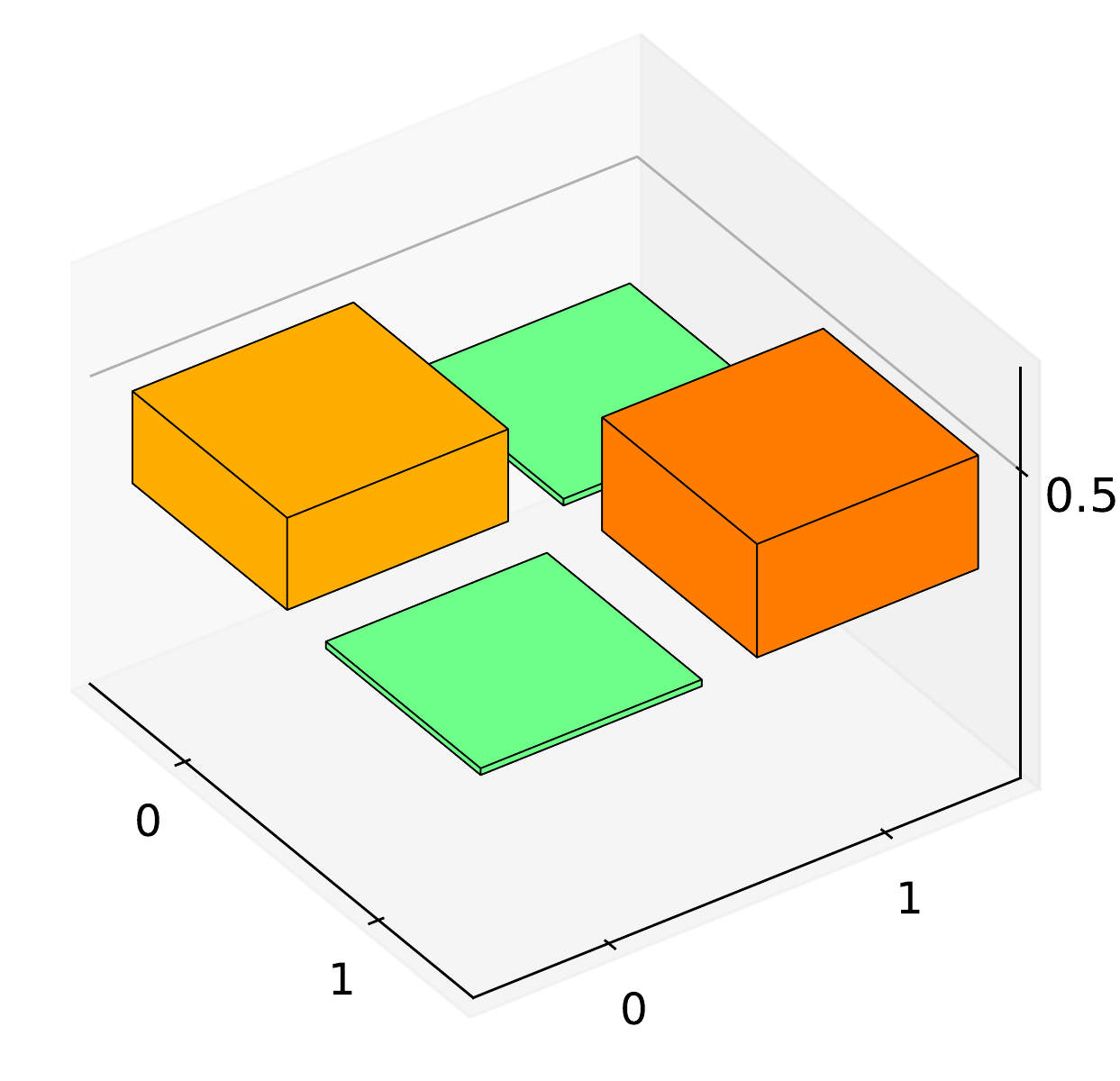}%
        \includegraphics[width=0.10\textwidth]{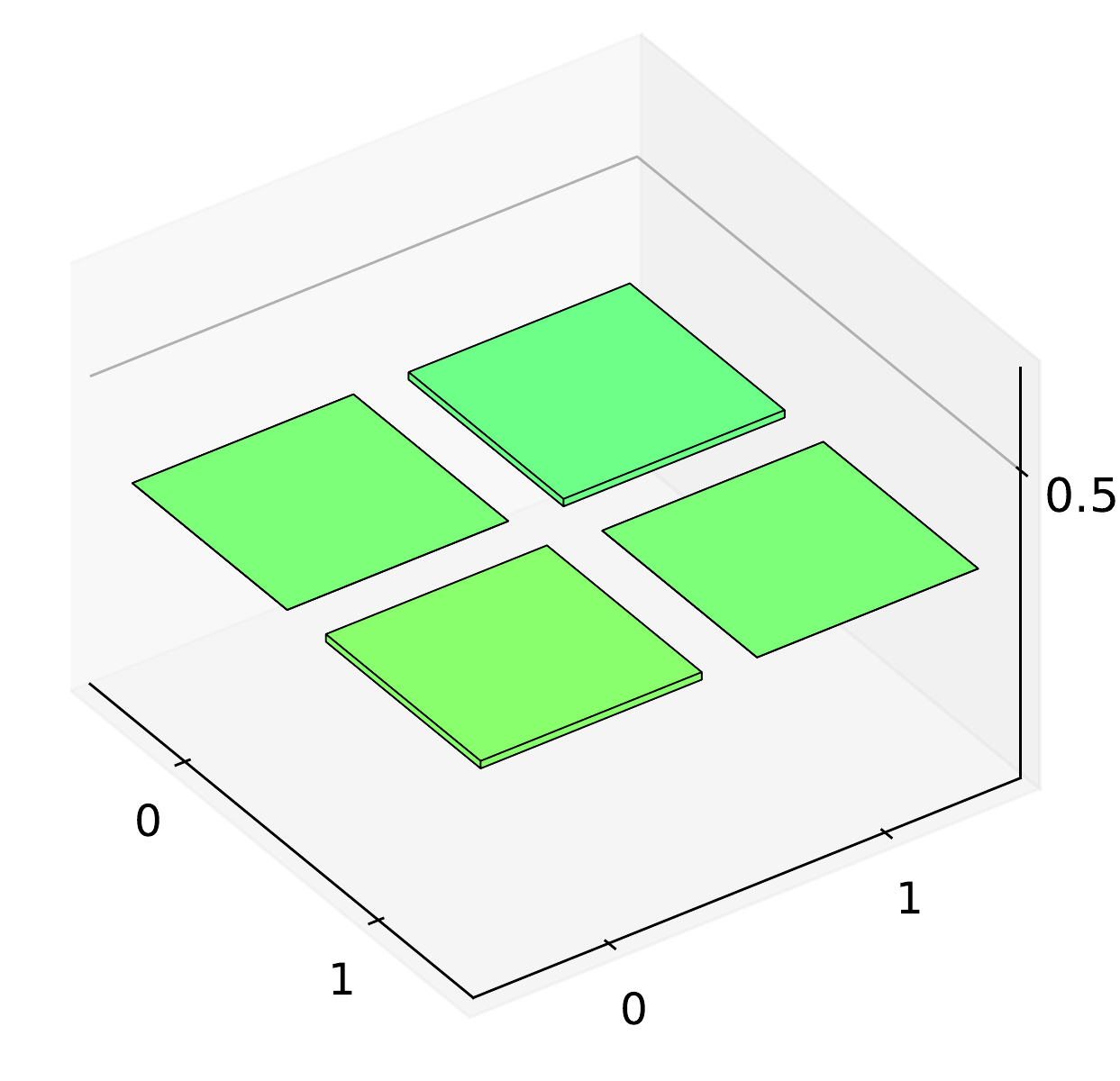}%
        \hfill%
        \includegraphics[width=0.10\textwidth]{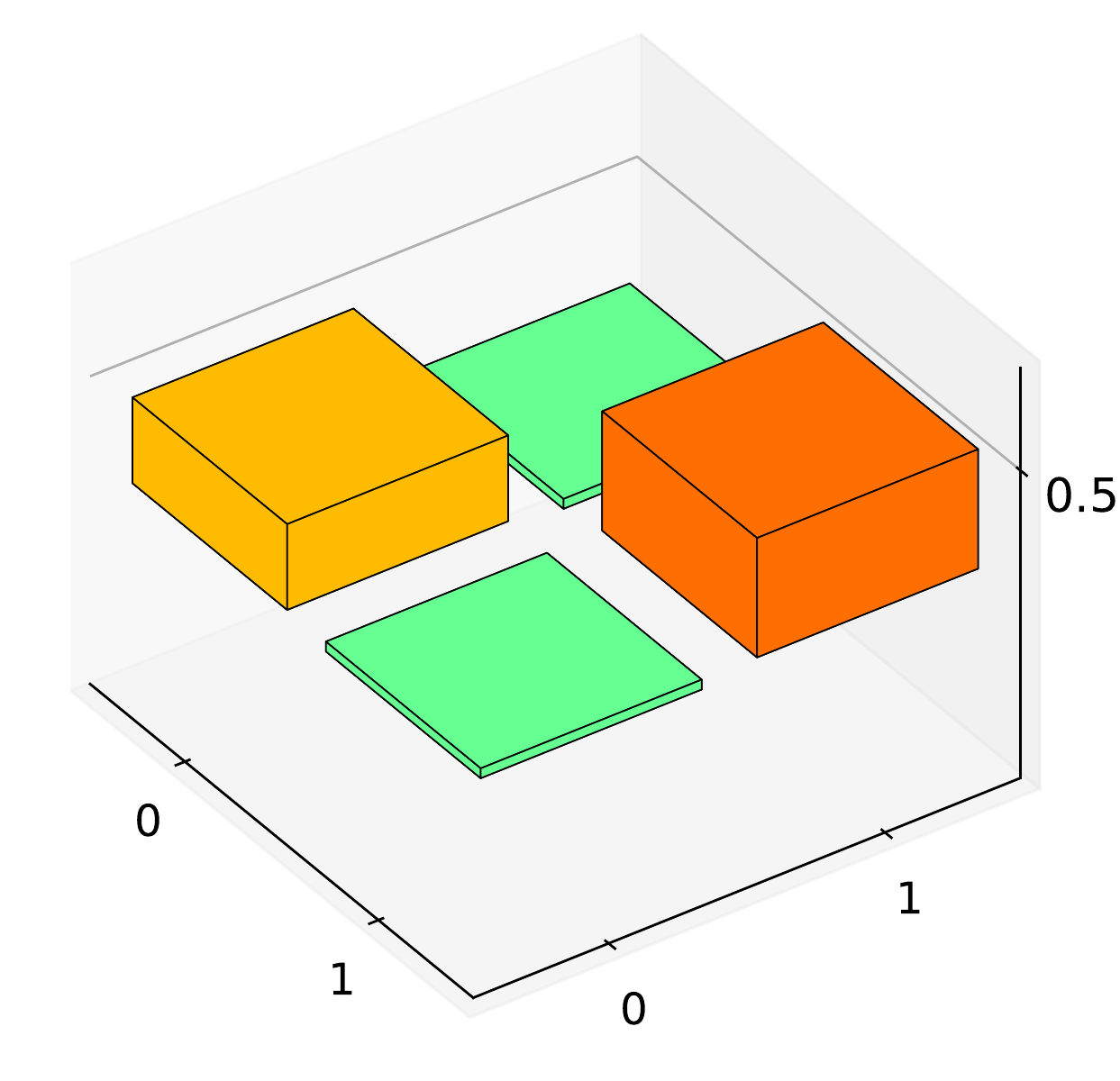}%
        \includegraphics[width=0.10\textwidth]{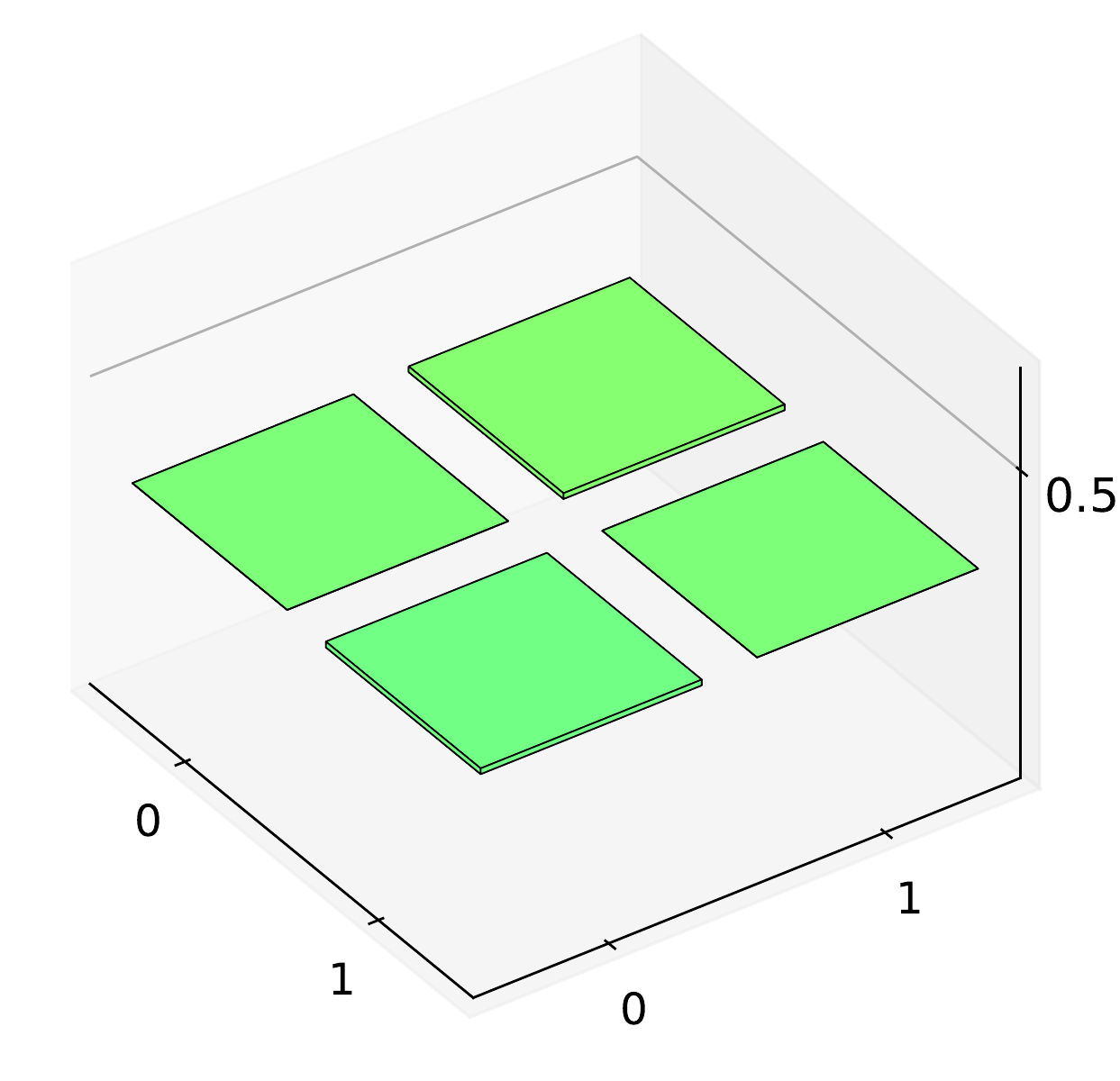}\\[1ex]
        \includegraphics[width=0.55\textwidth]{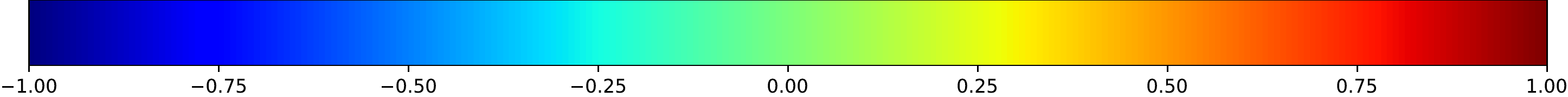}
    \caption{City state plots of density matrices for the pure quantum states (top row), ideal single qubit density matrices for $1 \rightarrow 9$ cloning (middle row), and the single qubit density matrices of clone index 0 from the H1-1 experiments (bottom row). The four columns correspond to the four message qubit states which are cloned. Each density matrix is plotted with two separate city state figures, one left hand plot for the real components of the density matrix and one right hand plot for the imaginary components of the density matrix. The magnitude of the density matrix scalar being plotted by each bar is color coded, the reference bar for which is given at the bottom of the figure. Plot y-axis scales are all in range $[-1, 1]$. }
    \label{fig:density_matrices}
\end{figure*}
%
\section{Results on Quantinuum H1-1}
\label{sec:results}
In this section the quantum telecloning results are shown for the $1 \rightarrow 9$ circuit applied to the four tetrahedral basis states. The circuits were executed on the Quantinuum H1-1 quantum computer. 

The left hand plot of Figure \ref{fig:result_barplots} shows the computed clone fidelities for all $M=9$ approximate clones. The clones achieve fidelities ranging from 0.52 to 0.67, which are below the theoretical optimum of $\frac{19}{27} \approx 0.7037$ as expected. The average fidelity (see Figure \ref{fig:results_averages}) across all states and clones is 0.59, thus falling near the middle of the range of $0.5$ (maximally mixed state) to $\frac{19}{27}$ (maximally achievable quantum information), thus indicating that some quantum information was retained.  

Figure \ref{fig:results_averages} shows the fidelity results averaged over the clone indices and the tetrahedral state indices.  Interestingly, there is a consistent trend where the $0$ state has higher clone fidelity compared the clones of the other 3 states. There is not a clear reason why the clone fidelities for the $0$ state are consistently better than the other states. A possible explanation could be due to non-uniform qubit decoherence \cite{single-qubit-state-tomography}, and it is relevant to note that similar state dependent trends were observed in the experimental results on an IBMQ device in \cite{pelofske2022telecloning}. Figure \ref{fig:results_averages} also shows a consistent decrease in clone fidelity as a function of clone index, with the only exception being clone index 0. We conjecture that this decrease is due to the number of two-qubit gates involved in computing the different clones.

The right hand plot of Figure \ref{fig:result_barplots} shows the proportions of the four different Bell state measurements made for each of the telecloning circuit executions of of $1500$ samples. The theoretical properties of the circuit are such that these proportions should all be $0.25$, and in these measurements observe reasonable consistency with this. 

Figure \ref{fig:density_matrices} shows the ideal density matrix representations of the pure message qubit state, the optimal approximate single qubit density matrix for $1 \rightarrow M$ cloning, and the reconstructed density matrices from the experimental results. The experimental density matrices are only from the index 0 clones, however since the telecloning circuit is symmetric these serve as a reasonable representation of what the clone reconstructions look like.

\section{Conclusion}
\label{sec:conclusion}

This article has presented an experimental demonstration of creating $9$ approximate symmetric clones of pure quantum states from a telecloning protocol that was executed on the Quantinuum H1-1 quantum computer, based on an algorithmic improvement for the circuit model construction. This experimental demonstration is the largest direct telecloning protocol that can be constructed and executed on current NISQ hardware as the H1-1 device is the largest current NISQ device (in terms of qubits) that has classical conditional gate operations. Other non-direct implementations could be constructed by using variants of the telecloning protocol which do not require mid-circuit measurement and classical conditional gates (for example post selection and deferred measurement \cite{pelofske2022telecloning}), which come at the cost of increased overhead in terms of circuit fidelity or number of shots. This experiment serves as a benchmark of the capabilities of quantum hardware; the clone fidelities show how effectively a reasonably deep (in terms of two qubit circuit depth) and wide (in terms of qubits) circuit can be executed while still retaining some of the quantum nature of the state. 

While we reported fidelity numbers as a main measure of how well the device performs, we should point out that trivial, non-optimal telecloning protocols \cite{scarani2005quantum} exist whose implementation even on a noise-free quantum device would only achieve fidelities below the theoretically achievable limits of Eq. \ref{eq:theoretical-fidelity}, but these somewhat aberrant protocols could practically match or outperform the measured fidelities. The difference to our optimal telecloning protocol is that it would achieve the theoretical bounds of Eq. \ref{eq:theoretical-fidelity} on an error-corrected device.

An open question with regards to constructing quantum telecloning states is whether there exists an algorithm for creating $1 \rightarrow M$ symmetric universal optimal telecloning (for $M \geq 4$) \emph{without} the need for ancilla qubits. Ancilla qubits restrict the size of the telecloning state that can be created on quantum hardware.


%
%

\bibliographystyle{plainurl}
\bibliography{references.bib}{}

\end{document}